      [1999/12/01 v1.4c Il Nuovo Cimento]
\documentclass{cimento}
\usepackage{graphicx}  
\begin{document}
\title{The JLab 12GeV Upgrade and the Initial Science
 Program} 
 \author{Volker D. Burkert}
\instlist{\inst{ Jefferson Lab, Newport News, VA 23606, USA }}
\PACSes{\PACSit{13.60.-r,13.60.Hb,13.60.Fz,13.60.Le,13.40.Gp,14.20.Gk}{}}
\maketitle
\vspace{-0.5cm}
\noindent \small{\it Lecture given at the International School of Physics
    'Enrico Fermi', 2011, Varenna, Italy}

\noindent

\vspace{2cm}
\begin{abstract}
An overview is presented of the upgrade of JLab's cw electron accelerator from a maximum beam energy of currently 6 GeV to 12 GeV. Construction of the upgrade project has started in 2008. A broad experimental program has been developed to map the nucleon's intrinsic correlated spin and momentum distribution through measurements of deeply exclusive and semi-inclusive processes, and to probe the quark and gluon confinement by studying the spectrum of mesons with exotic quantum numbers. Other programs include the forward parton distribution function at large $x_{B}$, the quark and gluon polarized distribution functions, the measurements of electromagnetic form factors of the nucleon ground state and of nucleon resonance transitions at high $Q^2$, and the exploration of physics beyond the Standard Model in high precision parity violating processes. The 12 GeV electron beam is also well suited to explore quark hadronization properties using the nucleus as a laboratory.
\end{abstract}
\vfill\eject
\noindent
{\bf{Table of Content}}\\\\\\
Page ~~~~~Section\\
\tableofcontents 
\vspace{2cm}
\section{Introduction}
\label{intro}

The challenge of understanding nucleon electromagnetic structure still 
continues after more than five decades of experimental scrutiny. From the initial 
measurements of elastic form factors to the accurate determination of 
parton distributions through deep inelastic scattering (DIS), the
experiments have increased in statistical and systematic accuracy.  It was
realized in recent years that the parton distribution functions
represent special cases of a more general, and much more powerful  way of 
characterizing the structure of the nucleon, the generalized parton 
distributions (GPDs)~\cite{Mueller:1994,Ji:1996nm,Ji:1996ek,Radyushkin:1996nd,Radyushkin:1997ki}.
  The GPDs describe the
simultaneous distribution of particles with respect to both position and 
momentum. In addition to the information about the spatial density (form factors) 
and momentum density (parton distribution), these functions reveal the 
correlation of the spatial and momentum distributions, {\it i.e.} how the 
spatial shape of the nucleon changes when probing quarks of 
different wavelengths. 

The concept of GPDs has led to completely new methods of ``spatial imaging''
of the nucleon, either in the form of two-dimensional tomographic images, or 
in the form of genuine 
three-dimensional images.  GPDs also allow us to 
quantify how the orbital motion of quarks in the nucleon contributes to the 
nucleon spin -- a question of crucial importance for our understanding of 
the ``dynamics'' underlying nucleon structure.  The spatial view of the 
nucleon enabled by the GPDs provides us with new ways to test dynamical 
models of nucleon structure. 
The mapping of the nucleon GPDs, and a detailed understanding of the
spatial quark and gluon structure of the nucleon, have been widely 
recognized as the key objectives of nuclear physics of the 
next decade. This requires a comprehensive program, combining results
of measurements of a variety of processes in electron--nucleon 
scattering with structural information obtained from theoretical studies, 
as well as with expected results from future lattice QCD simulations.
\begin{figure}[t]
\hspace{0.cm}
\resizebox{1.0\textwidth}{!}{
\includegraphics{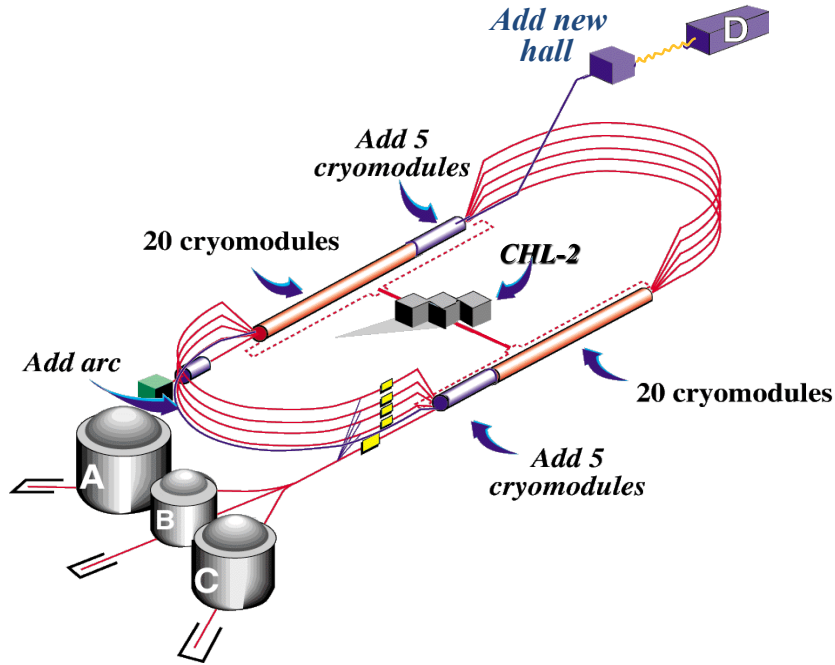}}
\vspace{-0.7cm}
\caption{The Jefferson Lab continuous electron beam accelerator facility showing the components needed for the 12 GeV upgrade.} 
\label{cebaf}    
\end{figure}
\begin{figure}[th]
\hspace{-0.90cm}
\resizebox{1.1\textwidth}{!}{
\includegraphics{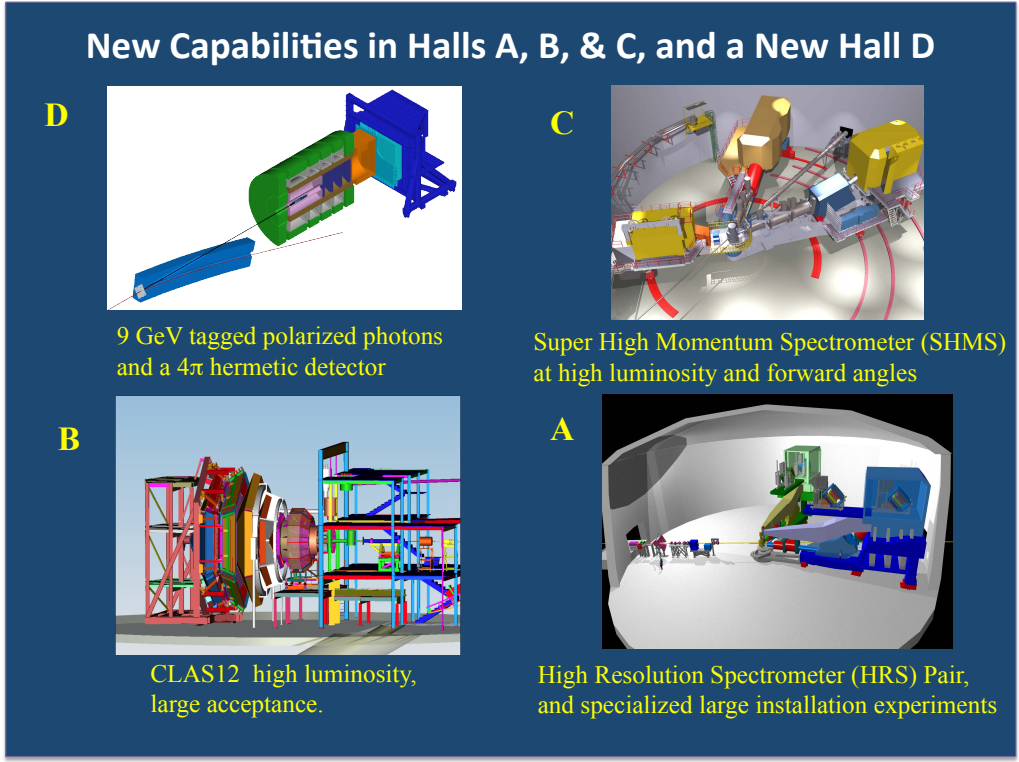}}   
\caption{Baseline experimental equipment to support the scientific program for the JLab 12 GeV
 energy upgrade Description in the text.} 
\label{equipment}
\end{figure}

While GPDs, and also the more recently introduced transverse momentum dependent 
distribution functions (TMDs), open up new avenues of research, the
traditional means of studying the nucleon structure through 
electromagnetic elastic and transition 
form factors, and through flavor- and spin-dependent parton distributions must also be 
employed with high precision to extract physics on the nucleon structure in the transition 
from the regime of quark confinement to the domain of asymptotic freedom. These 
avenues of research can be explored using the 12 GeV cw beam of the JLab 
upgrade with much higher precision than has been achieved before, 
and can help reveal some of the remaining secrets 
of QCD. Also, the high luminosity available 
will allow us to explore the regime of extreme 
quark momentum, where a single quark carries 80\% or more of the proton's 
total momentum. 

The strong force is mediated by the exchange of gluons between quarks as well as between 
gluons. Their presence has been observed in high energy scattering experiments where 
they generate a third jet of particles in addition to the 2 jets coming from the fragmentation
of the scattered quarks. A clear understanding of the role of gluons in interactions at at the 
lower energies is still missing. Models that describe the interaction through gluon flux tubes predict
that the glue plays an active role and can be excited and lead to novel states of mesons and baryons.
The search of some of these hybrid states is the focus of the photo production experiments in 
the new Hall D.      
The plan for this report is to give a brief overview in section~\ref{equip} of the upgrade of the JLab accelerator to 12 GeV 
and briefly discuss the experimental equipment 
that is currently under construction to cary out the science program during the first 5 years of 
operation. In section~\ref{gpds} and section~\ref{tmds} I discuss the program to study the generalized
parton distribution function (GPDs) and the transverse momentum dependent functions (TMDs) 
using polarization degrees-of-freedom, which covers a major part of this contribution and are the main
ingredients to develop the multi-dimensional representation of quarks in the nucleon in coordinate and 
momentum space. In  sections~{\ref{spectroscopy} through~\ref{new-physics}, I discuss some 
of the other programs: spectroscopy, form factors and structure functions at high $Q^2$, nuclear processes, 
and physics beyond the SM. Since this is an overview covering a broad future physics program, 
I will not be able to go into depth of any of the topics discussed. The phenomenological 
underpinnings of much of the material, especially the part related to GPDs and TMDs, has been 
presented in lectures and seminars in the earlier part of the school by experts, and I refer the 
reader to these contributions for details in this volume.     

\begin{table}[ht]
\caption{\rm Sensitivity of deeply virtual Compton scattering and deeply virtual meson production processes to the leading twist GPDs for different quark flavors. }
\resizebox{1.0\textwidth}{!}{%
 \includegraphics{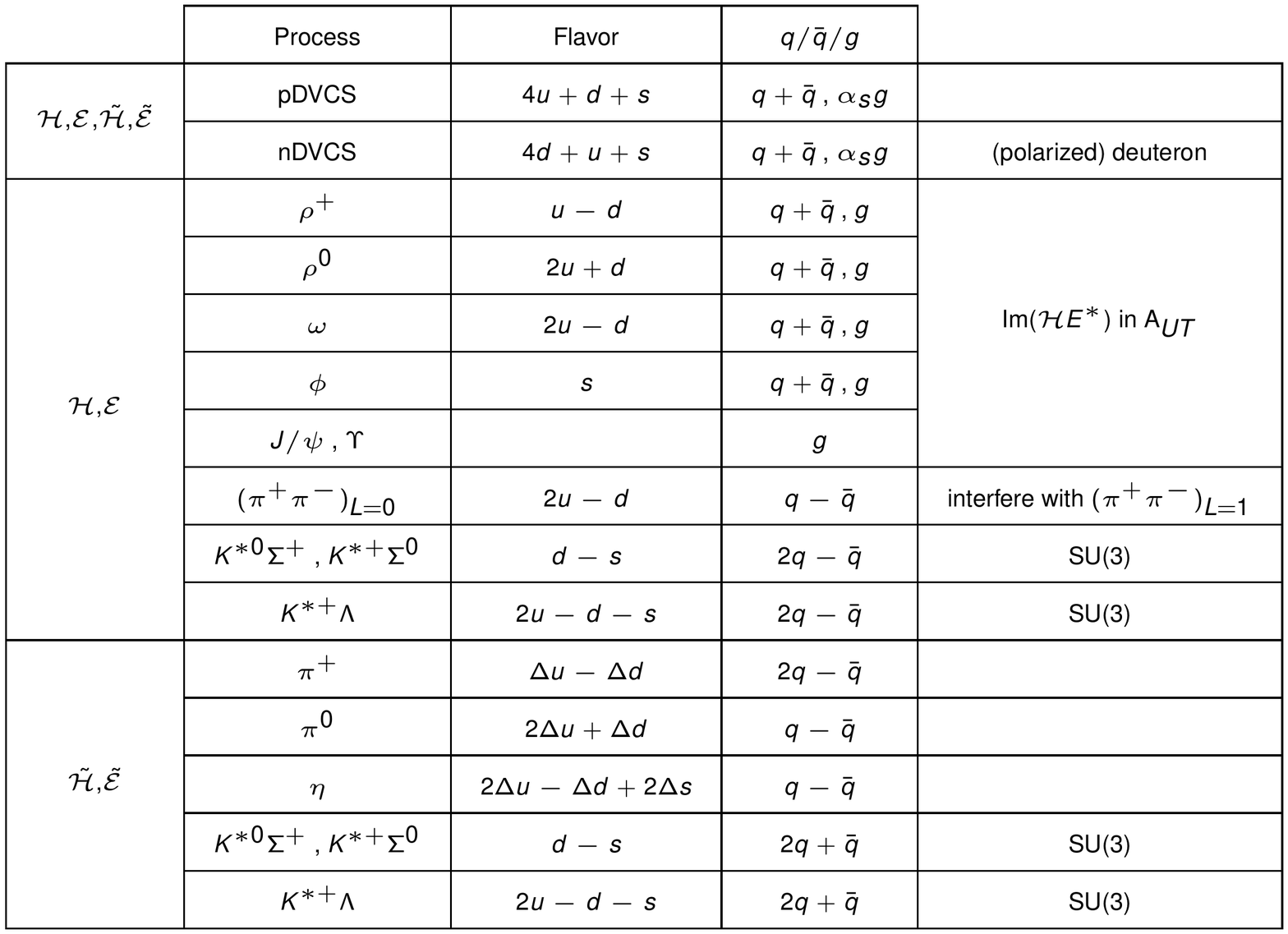}}
\label{access_gpds}    
\end{table}

\section{The Electron Accelerator and the 12 GeV Experimental Equipment. } 
\label{equip}
\noindent The electron accelerator is shown schematically in Fig.~\ref{cebaf}. The two linear accelerators are based on superconducting rf technology. Spin polarized electrons are generated in the gun and pre-accelerated to 50MeV in the injector shown at the upper left end of the racetrack. They are then boosted in the north linac to about 600 MeV.  They are then bent by 180 degrees and injected into the south linac to be accelerated to up to 1200 MeV. This is repeated four more times when the final energy of 6000 MeV is
reached. For the energy upgrade five accelerating cryo-modules with four times higher gradients per unit length are added to both linacs to reach a maximum energy at the existing end stations of nearly 11 GeV. One arc and one more path through the north linac are added to accelerate the beam to 12 GeV and transport it to the new Hall D. 
Major equipment upgrades are planned to support a broad program of nuclear physics, hadron spectroscopy, and the exploration of the quark-gluon dynamics and multi-dimensional structure of the nucleon. The base equipment that is part of the 12GeV upgrade scope is under construction and is scheduled for completion in the fall of 2014. The base 
equipment in the four Halls is shown in Fig.~\ref{equipment}.
The focus in Hall D is on the study of hybrid mesons with exotic quantum numbers to understand the excitation of the glue in hadronic systems. Hall D houses a large hermitic detector that is based on a solenoid magnet, and has tracking capabilities and photon detection over nearly the full $4\pi$ solid angle. The bottom left panel shows the new {\tt CLAS12} large acceptance spectrometer, with large angular coverage and increased luminosity. The top panel on the right shows the spectrometer pair of forward focusing spectrometers in Hall C including the new SHMS spectrometer on the left. In Hall A the existing HRS2 spectrometer pair will be retained. In addition, the Hall will house large installation equipment for specialized experiments. The solenoid detector in Hall D, {\tt CLAS12} and the SHMS are part of thr base equipment, and are projected to be available for physics by the end of 2014. There are additional pieces of equipment proposed that are outside the scope of the 12GeV project and require separate funding. In Hall A several proposals are under consideration, a super gig bite spectrometer (SBS) with large solid angle, a spectrometer (Moller) to measure parity violation in electron-electron scattering, and a solenoid detector for parity violation experiments in DIS and for SIDIS experiments (SOLID). There are also efforts underway to incorporate a large acceptance RICH detector into {\tt CLAS12} for improved charged particle identification capabilities, and various ancillary detectors for neutron detection and quasi-real photon tagging. Complementing the new equipment that is under construction, are the highly polarized electron gun, high power cryogenic targets, and several polarized targets using  NH$_3$, ND$_3$, HD, butanol, and $^3$He to support a broad range of polarization measurements.
\section{Generalized Parton Distributions and DVCS}
\label{gpds}
\begin{figure}[ht]
\resizebox{1.0\textwidth}{!}{%
\includegraphics{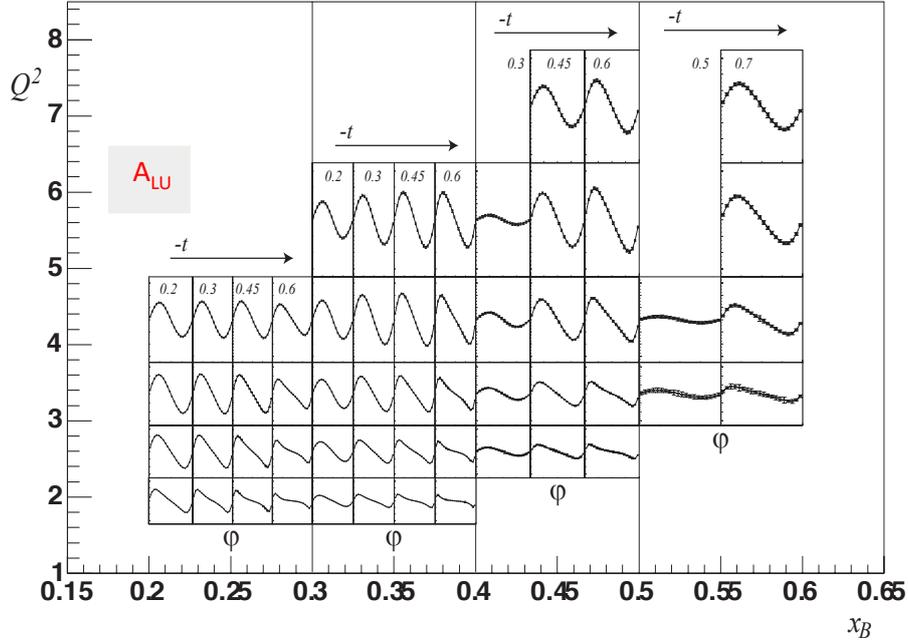}}
\caption{Projected data for the azimuthal dependence of the beam spin asymmetry $A_{LU}$ for 
 experiment ~\cite{e12-06-119}. }
\label{ALU_DVCS}    
\end{figure}
\begin{figure}[t]
\resizebox{1.0\textwidth}{!}{%
\includegraphics{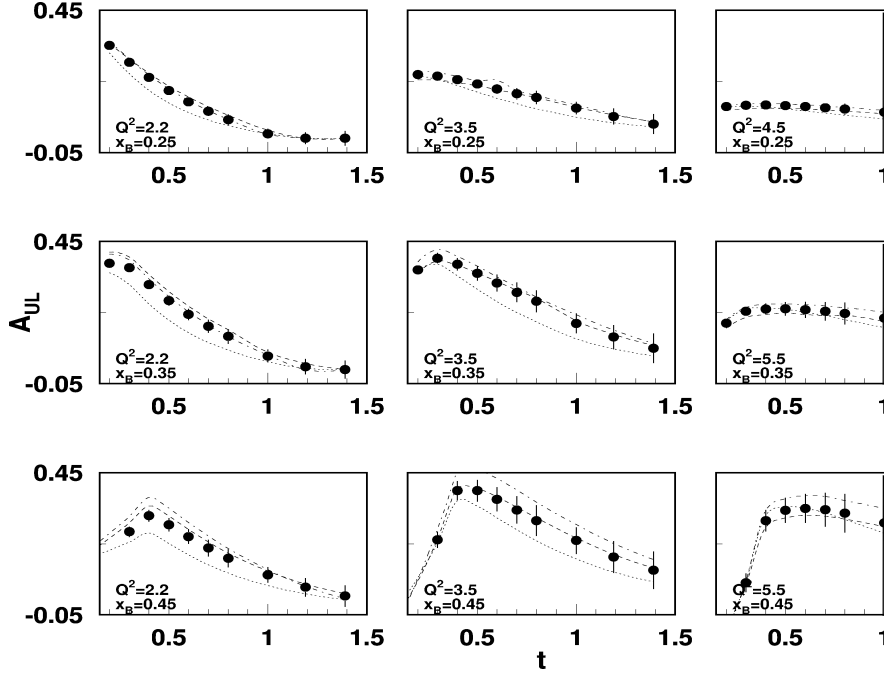}}
\caption{Projected data for the t-dependence of the longitudinal target spin asymmetry $A_{UL}$  for 
 experiment ~\cite{e12-06-119}. }
\label{AUL_DVCS}    
\end{figure}

It is now well recognized~\cite{Ji:1996nm,Belitsky:2001ns,Burkardt:2002hr,Belitsky2004} that 
deeply virtual Compton scattering and deeply virtual meson production are most suitable for
mapping out the twist-2 vector GPDs $H,~E$ and the axial GPDs ${\tilde H},~{\tilde E}$ in $x,~\xi,~t$, 
where $x$ is the momentum fraction of the 
struck quark, $\xi$ the longitudinal momentum transfer to the quark, and $t$ the 
transverse momentum transfer to the nucleon. Having access to a 3-dimensional image
of the nucleon (two dimensions in transverse space, one dimension in longitudinal 
momentum) opens up completely new insights into the complex internal structure and dynamics 
of the nucleon. In addition, GPDs carry information of more global nature. For example,
the nucleon matrix element of the energy-momentum tensor 
contains 3 form factors that encode information on the angular 
momentum distribution $J^q(t)$ of the quarks with flavor $q$ in transverse space, their 
mass-energy distribution $M_2^q(t)$, and their pressure and force 
distribution $d^q_1(t)$. How can we access these form factors? The only 
known process to directly measure them is elastic graviton scattering 
off the nucleon. However, these form factors also 
appear as moments of the two vector GPDs~\cite{goeke2007}, thus offering
prospects of accessing gravitational form factors through the detailed mapping of GPDs.  The 
quark angular momentum form factor in the nucleon is given by 
\begin{figure}[h]
\hspace{-1.5cm}
\resizebox{1.1\textwidth}{!}{%
\includegraphics{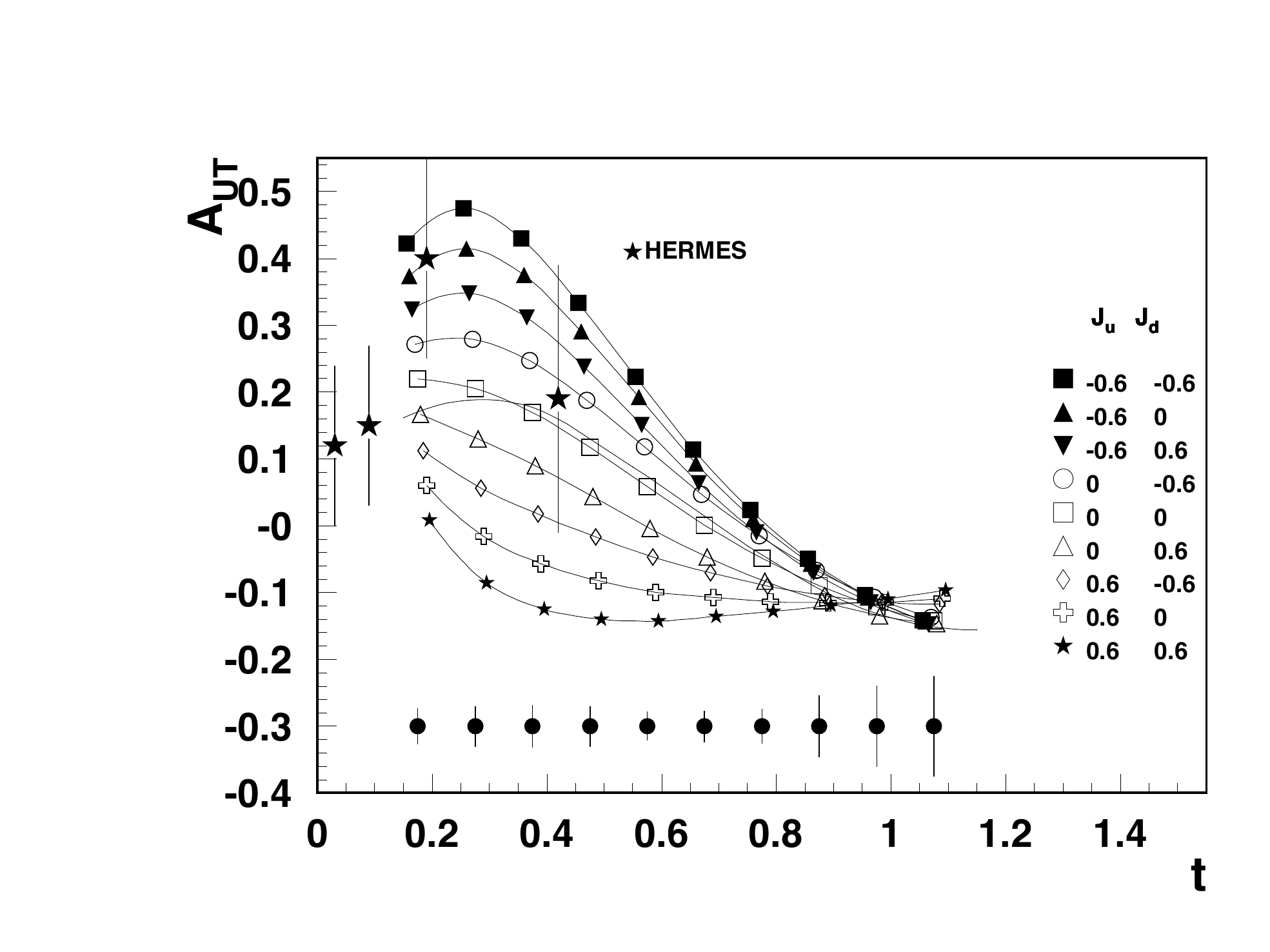}}
\caption{Projected data for the t-dependence of the transverse target spin asymmetry 
$A_{UT}^{\sin{\phi}\cos{\phi-\phi_s}}$. The $t$ dependence of $A_{UT}$ is 
 shown for a single bin at $Q^2=2$GeV$^2$ and $x$=0.25~\cite{loi11-105}.}
\label{AUT_DVCS}  
\end{figure}
$$J^q(t) = 
\int_{-1}^{+1}dx x [H^q(x, \xi, t) + E^q(x, \xi, t)]~,$$ and the mass-energy and pressure 
distribution $$M_2^q(t) + 4/5d^q_1(t)\xi^2 
= \int_{-1}^{+1}dx x H^q(x, \xi, t)~.$$ The mass-energy and force-pressure distribution 
of the quarks are given by the second moment of GPD $\it{H}$, and their relative 
contribution is controlled by $\xi$. A separation of $M^q_2(t)$ and 
$d^q_1(t)$ requires measurement of these moments in a large range of 
$\xi$. The beam helicity-dependent cross section asymmetry is given 
in leading twist as 
$$ A_{LU} \approx \sin\phi[F_1(t)H + \xi(F_1+F_2)\tilde{H}]d\phi~, $$where
$\phi$ is the azimuthal angle between the electron scattering plane and the hadronic production
plane. The kinematically suppressed term with GPD $E$ is omitted. 
The asymmetry is mostly sensitive to the GPD $H(x=\xi,\xi,t)$. In a wide kinematics~\cite{clas-dvcs-1,clas-dvcs-3}  
the beam asymmetry $A_{LU}$ was measured at Jefferson Lab at modestly high $Q^2$, $\xi$, and $t$, and in a more limited kinematics~\cite{halla-dvcs} the cross section difference
$\Delta\sigma_{LU}$ was measured with high statistics. Moreover, 
a first measurement of the target asymmetry 
$A_{UL}=\Delta\sigma_{UL}/2\sigma$ was carried out~\cite{clas-dvcs-2}, where 
$$A_{UL} \approx \sin\phi[F_1\tilde{H} + \xi(F_1+F_2)H]~.$$  
The combination of $A_{LU}$ and $A_{UL}$ allows the separation of GPD $H(x=\xi,\xi,t)$ and
$\tilde{H}(x=\xi,\xi,t)$.  
Using a transversely polarized target the asymmetry 
$$A_{UT} \approx \cos\phi\sin(\phi-\phi_s) [t/4M^2 (F_2H - F_1 E)] $$ can be measured, 
where $\phi_s$ is the azimuthal angle of the target polarization vector relative to the 
electron scattering plane. $A_{UT}$ depends in leading order on GPD $E$. Measurement of $A_{UT}$ may thus
be the most efficient way of determining GPD $E$. 
\begin{figure}[ht]
\hspace{1.0cm}
\resizebox{0.85\textwidth}{!}{%
  \includegraphics{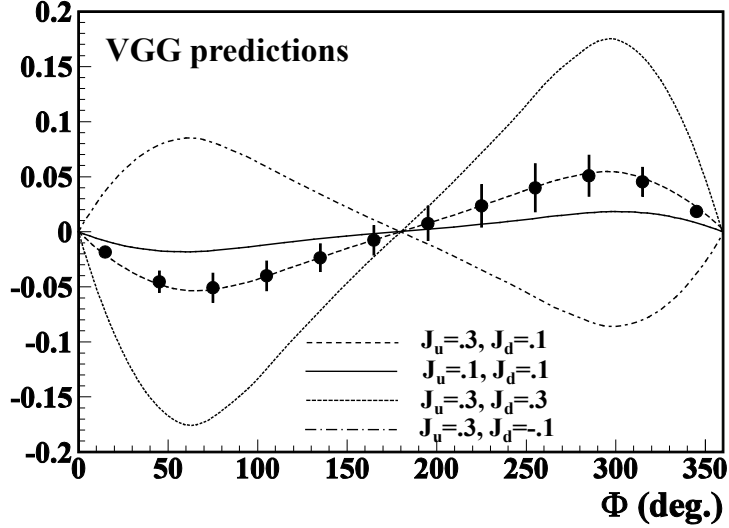}}
\caption{Projections of experiment~\cite{e12-11-003} for the DVCS beam-spin asymmetry $A_{LU}$ on the neutron for
one bin in $Q^2$, $x_B$, $t$. The different lines indicate expected asymmetries for different values of the nucleon spin carried by $u$ and $d$ quarks as predicted by the VGG model~\cite{ vgg}. 
The magnitude of $A_{UL}$ on the neutron is predicted much smaller than what has been measured on the proton. }
\label{neutron-dvcs}    
\end{figure}  
  
\begin{figure}[t]
\hspace{-1cm}
\resizebox{1.1\textwidth}{!}{%
  \includegraphics{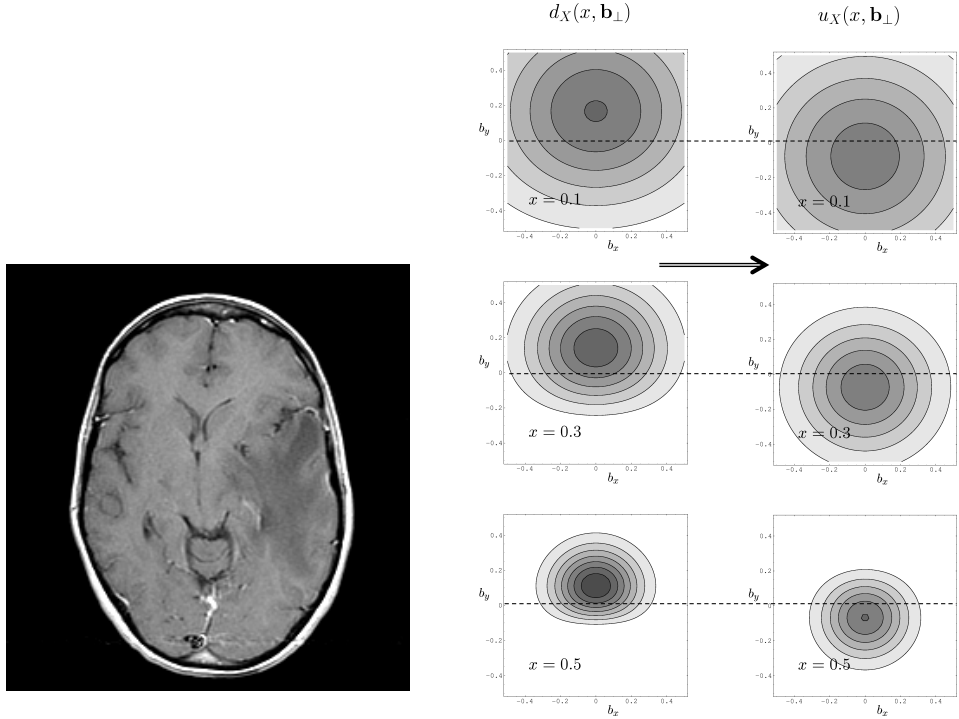}}
\caption{Left panel: Slice of a 3-D tomographic image of a human brain from an CAT scan. Right panel: Simulated quark distributions in a transversely polarized proton. The arrow indicates the direction of the polarization. 
The centroids of the d-quarks (left column) appear shifted upwards and the distributions appear deformed from the symmetric distribution in unpolarized protons; the centroids of the u-quark distributions are shifted in the opposite direction, generating a $u-d$ quark flavor dipole.  }
\label{images}    
\end{figure}  
First DVCS experiments carried out at JLab~\cite{clas-dvcs-1,clas-dvcs-3,halla-dvcs,clas-dvcs-2} and DESY~\cite{hermes-dvcs} showed promising 
results in terms of the applicability of the handbag mechanism to probe GPDs. The 12 GeV upgrade offers much improved possibilities for accessing GPDs. 
Figure~\ref{ALU_DVCS} shows the expected statistical precision of 
the DVCS-BH beam asymmetry for some sample kinematics. Using a
longitudinally polarized target one can also measure 
the  target spin asymmetries $A_{UL}$ with high precision. Figure~\ref{AUL_DVCS} shows 
the expected statistical accuracy for the moment $A_{UL}^{sin\phi}$. Using a transversely polarized 
target, the transverse target spin asymmetry $A_{UT}$ can be measured.  The $t-$dependence of
 $A_{UT}$ as projected in~\cite{loi11-105}  is shown in Fig.~\ref{AUT_DVCS} for a single bin in 
 $Q^2$ and $x$.  
 
 Measurements of all 3 asymmetries will allow a separate determination of GPDs 
$H,~\tilde{H}$ and $E$ at the above specified kinematics. If the $t$-dependences are known, 
 a Fourier transformation of GPD $E(x=\xi,t)$ can be used to determine the polarized quark distribution 
in transverse impact parameter space. To obtain a complete picture of the quark distribution in the nucleon, 
the GPDs need to be measured for both flavors $u$-quarks and $d$-quarks. 
This requires measurement of DVCS not only on the proton but also on the neutron. 
Experiment~\cite{e12-11-003} will measure the beam-spin asymmetry for the neutron 
$e n (p_s) \to e \gamma n (p_s)$

Figure~\ref{images} shows simulated images of the quark distributions in transverse space using a
model parameterization of GPDs~\cite{Burkardt:2002hr}.

Deeply virtual exclusive meson production (pseudo scalar mesons and vector mesons) will play an 
important role in disentangling 
the flavor- and spin-dependence of GPDs (see  table~\ref{access_gpds}). For exclusive mesons 
only the longitudinal 
photon coupling in $\gamma^* p \rightarrow Nm$ enables direct access to GPDs through 
the handbag mechanism and must be isolated from the transverse coupling. That the transverse
contribution cannot be neglected at currently available energies of 6 GeV was observed in 
the non-zero beam asymmetry measured with CLAS~\cite{demasi} that indicated the presence of 
a significant longitudinal-transverse interference term in the amplitudes. In addition, 
the dominance of the handbag mechanism in the longitudinal cross section must first be 
established at the upgrade energy before meson production may be incorporated in a 
global fit of all hadronic reactions to extract GPDs. 

\section{Semi-inclusive DIS and TMDs}
\label{tmds}
\noindent 
Semi-inclusive deeply inelastic scattering (SIDIS) processes, where the leading, 
high momentum hadron is detected in 
coincidence with the scattered lepton, are used for ``flavor tagging'' of quarks to select  
contributions from different quark species. Currently, the emphasis is to study SIDIS 
processes  that encode information on the 
transverse momentum distributions of quarks, information that is not otherwise accessible. 
For example, azimuthal distributions of final state particles in SIDIS  
provide access to the orbital motion of quarks and 
play an important role in the study of transverse-momentum dependent distributions (TMDs) 
of quarks in the nucleon. 
\begin{table}[tb]
\caption{{{\small Leading-twist transverse momentum-dependent distribution 
functions.  $U$, $L$, and $T$ stand for transitions of unpolarized, 
longitudinally polarized, and transversely polarized nucleons (rows) to 
corresponding quarks (columns).}}\label{tab1}} 
\begin{tabular}{|c|c|c|c|} \hline
N/q & U & L & T \\ \hline
 {U} & ${\bf f_1}$   & & ${ h_{1}^\perp}$ \\ \hline
 {L} & &${\bf g_1}$ &    ${ h_{1L}^\perp}$ \\ \hline
 {T} & ${ f_{1T}^\perp} $ &  ${ g_{1T}}$ &  ${ \bf h_1}$ \, ${ h_{1T}^\perp }$ \\
\hline
\end{tabular}
\end{table}
\begin{figure}[bht]
\resizebox{0.8\textwidth}{!}{%
 \includegraphics{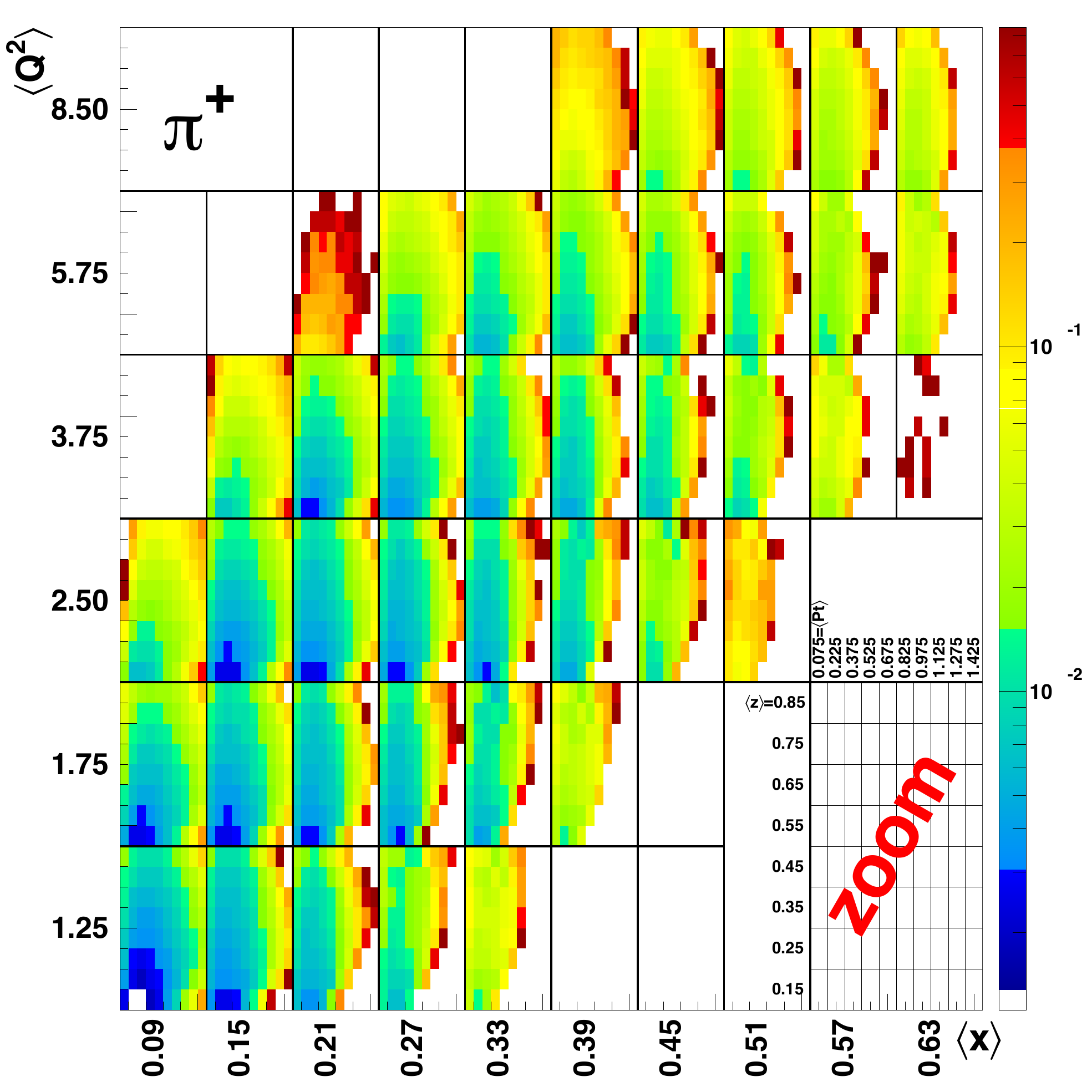}}
\caption{Kinematic coverage and projected uncertainties of the 4-dimensional $\pi^+$ yield 
in $Q^2$, $x$, $z$, and $p_T$. The color code shows the relative errors in each bin.}
\label{sidis}    
\end{figure}

TMD distributions describe transitions of a nucleon 
with one polarization in the initial state to a quark with another polarization 
in the final state. The diagonal elements in the table~\ref{tab1} are the momentum, 
longitudinal and transverse spin distributions of partons, and represent well-known 
parton distribution functions related to the square of the leading-twist, light-cone 
wave functions. Off-diagonal elements require non-zero orbital angular 
momentum and are related to the wave function overlap of $L$=0 and $L$=1 Fock 
states of the nucleon.  The chiral-even distributions 
$f_{1T}^\perp$ and $g_{1T}$ are the imaginary parts of the corresponding
interference terms, and the chiral-odd $h_1^\perp$ and $h_{1L}$ are the
real parts.  The TMDs $f_{1T}^\perp$ and  $h_{1}^\perp$ are related to 
the imaginary part of the interference of wave functions for different orbital 
momentum states and are known as the Sivers~\cite{sivers} and 
Boer-Mulders~\cite{boer,mulders} functions. They describe unpolarized quarks in the 
transversely polarized nucleon and transversely polarized quarks in the 
unpolarized nucleon, respectively.  
The most simple mechanism that can lead to a Boer-Mulders function is a 
correlation between the spin of the 
quarks and their orbital angular momentum. In combination with a final state 
interaction that is on average attractive, already a measurement of the sign 
of the Boer-Mulders function, would thus reveal the correlation between 
orbital angular momentum and spin of the quarks. 
\begin{figure}[tb]
\hspace{-0.7cm}
\resizebox{0.55\textwidth}{!}{%
	\includegraphics{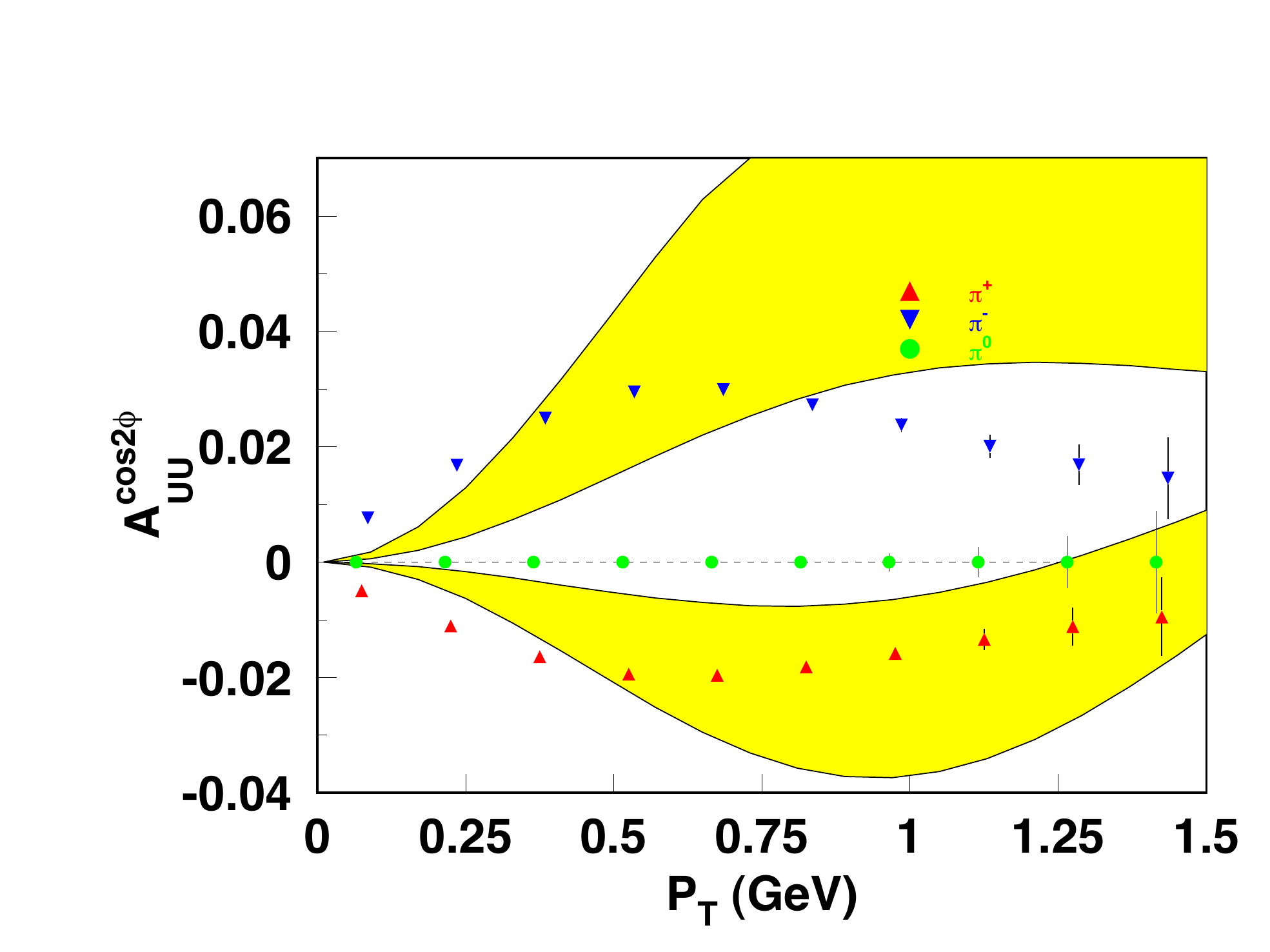}}
	\hspace{-1.0cm}
	\resizebox{0.55\textwidth}{!}{%
	\includegraphics{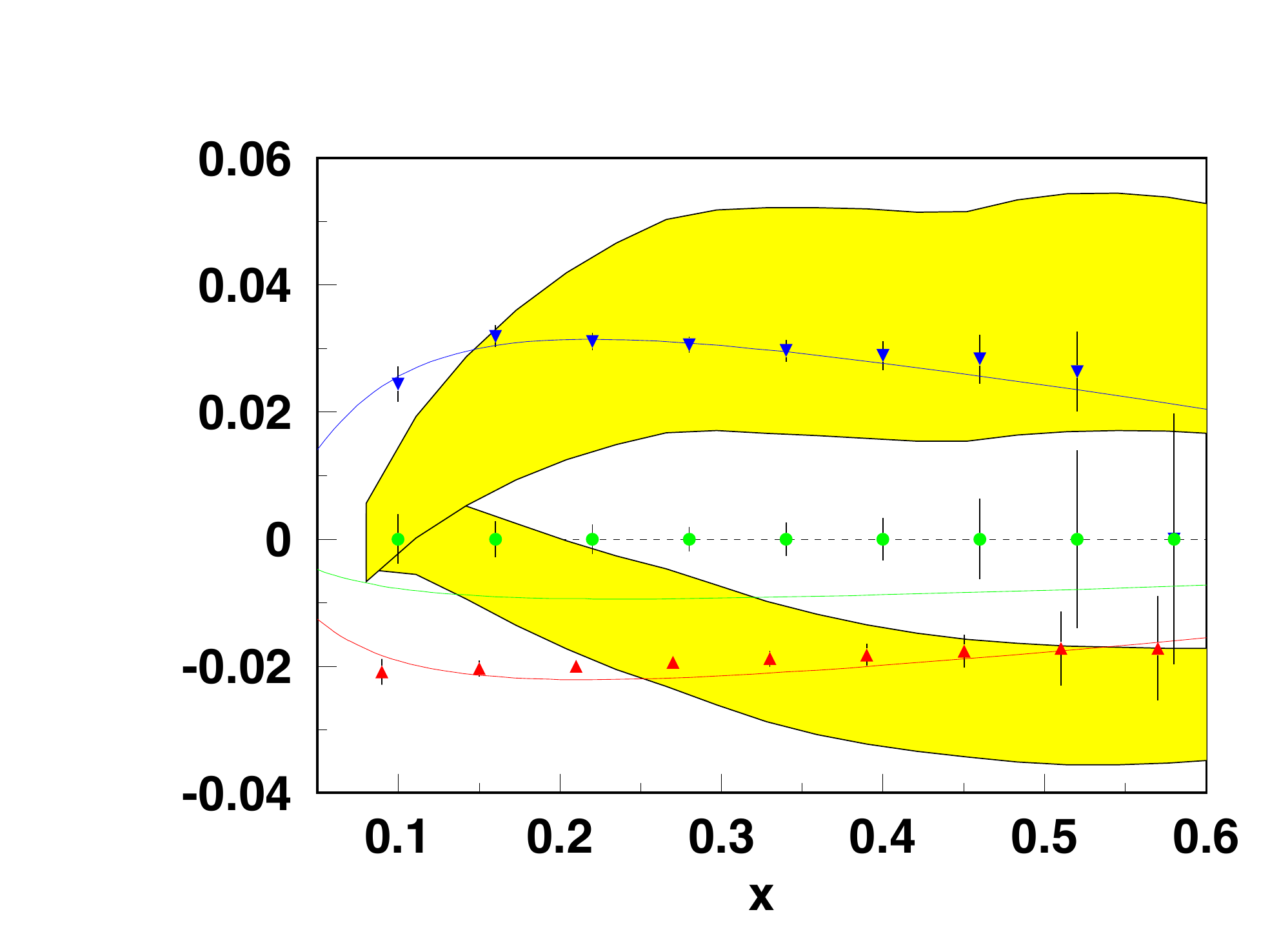}}
\caption{The $\cos2\phi$ moment (Boer-Mulders asymmetry) for pions
as a function of $Q^2$ and $P_T$ for $Q^2>2$~GeV$^2$ with {\tt CLAS12} 
at 11~GeV from 2000~hours of running.  Values are calculated assuming
$H_1^{\perp u\rightarrow \pi^+}=-H_1^{\perp u\rightarrow \pi^-}$. Only statistical
uncertainties are shown. The yellow band between the two curves indicates the range of two model predictions.}
\label{fig:Boer-Mulders}    
\end{figure}   
\begin{figure}[bt]
\vspace{-4cm}
\hspace{1cm}
 \resizebox{0.8\textwidth}{!}{%
 \includegraphics{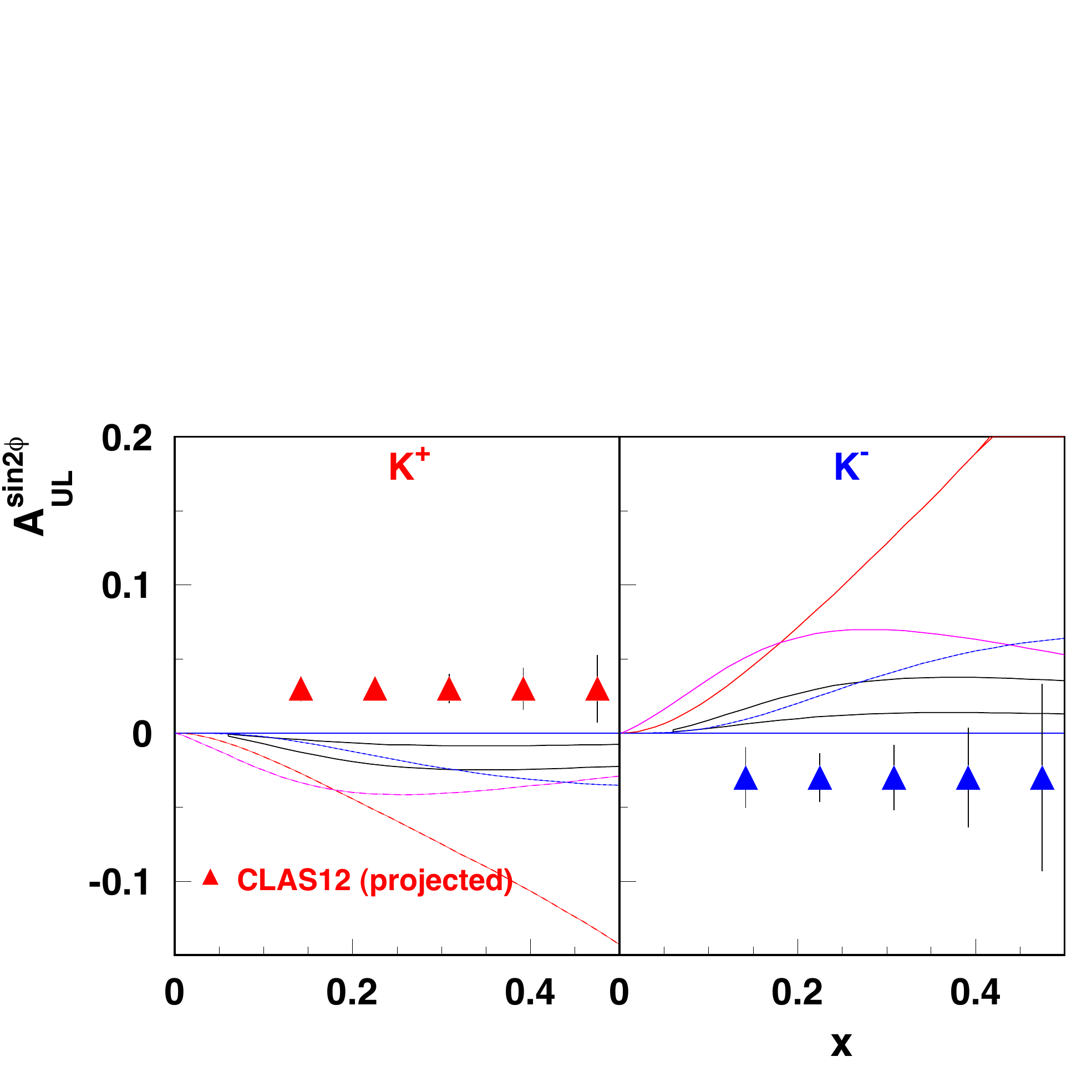}}
\caption{Projected x-dependence of the polarized spin target asymmetry at 11 GeV 
with {\tt CLAS12} for kaons for one bin in $Q^2,~z,~p_T$.}
\label{sidis-kaon-AUL}    
\end{figure}

\begin{figure}[h]
\vspace{0.0cm}
\resizebox{0.5\textwidth}{!}{%
 \includegraphics{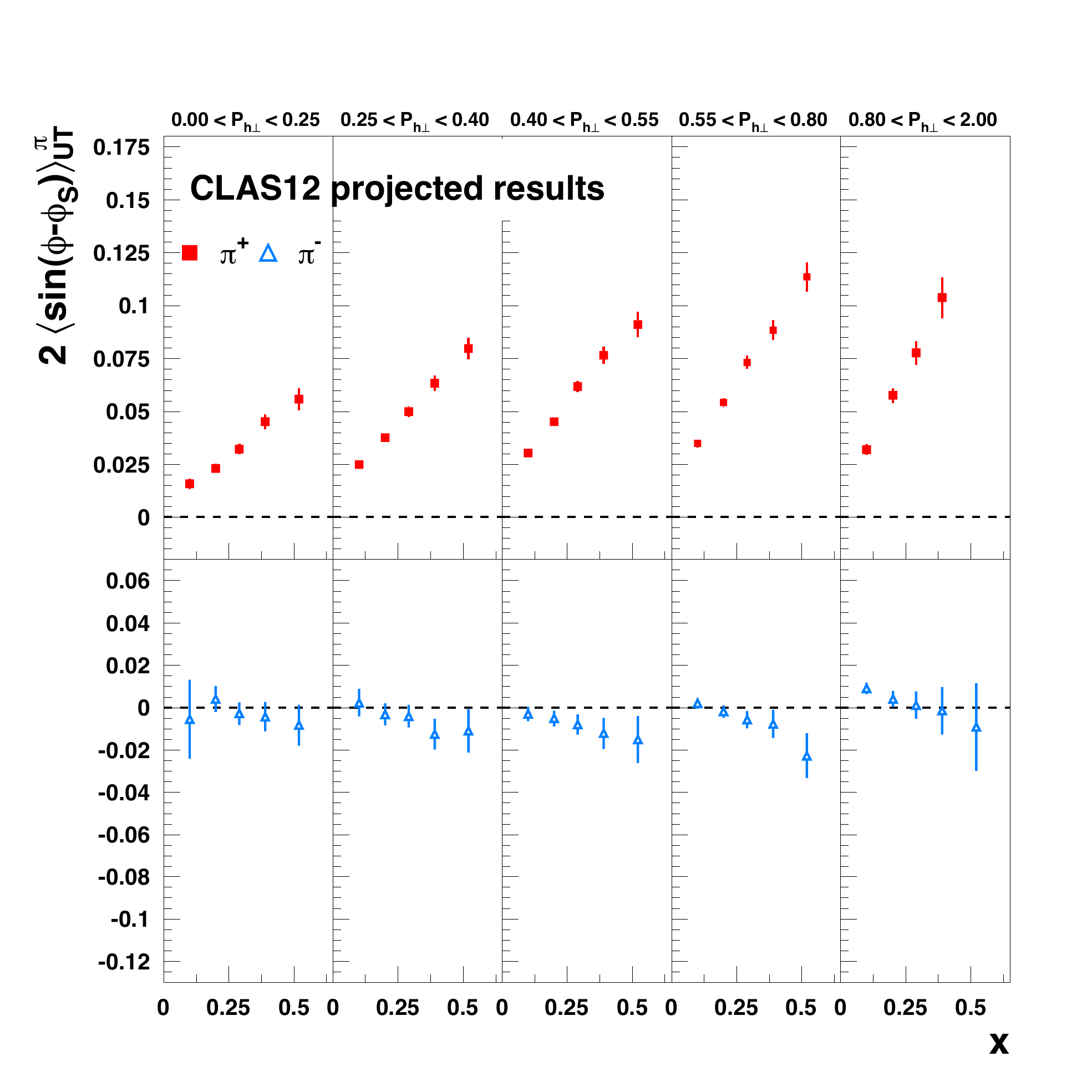}}
\resizebox{0.5\textwidth}{!}{%
 \includegraphics{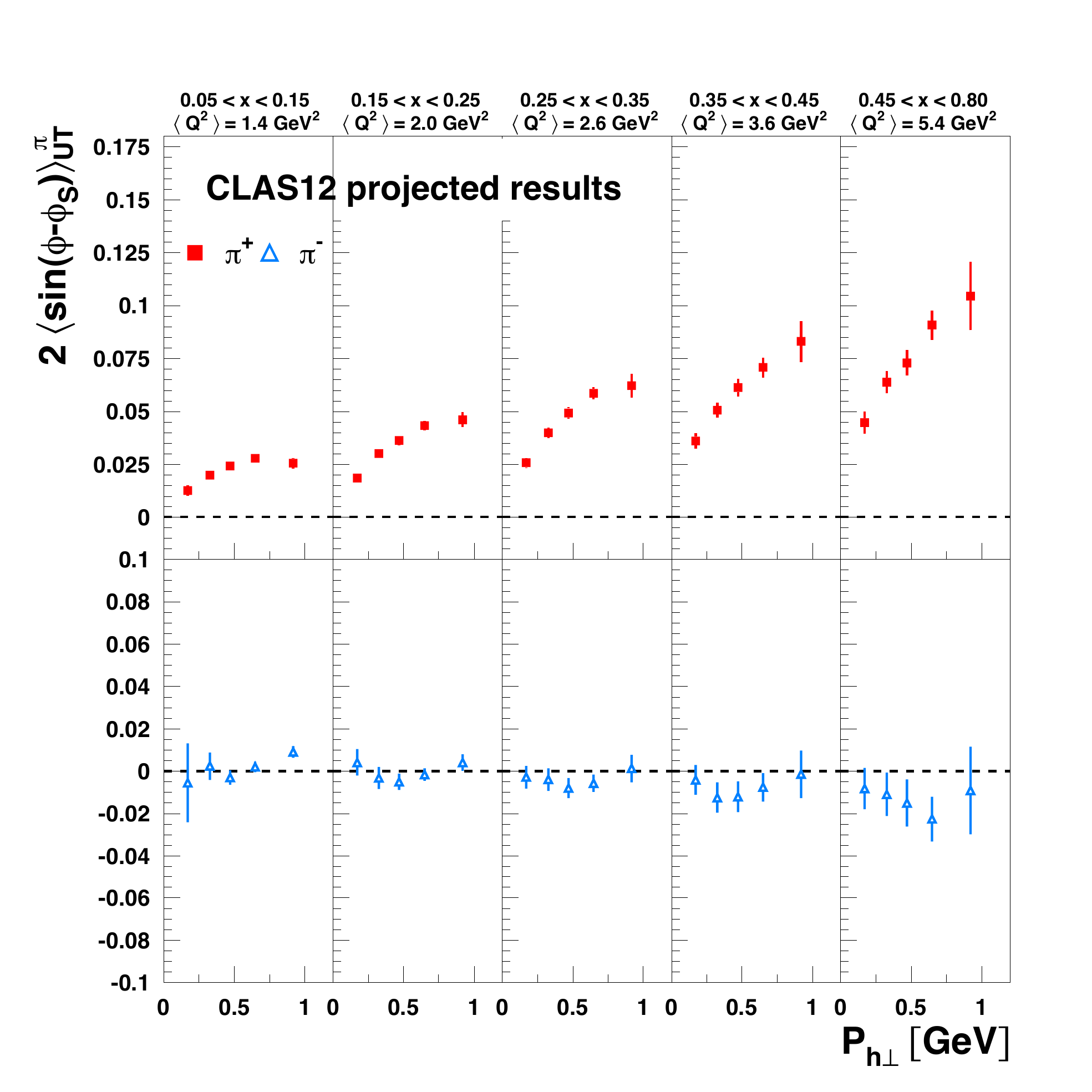}}
 \vspace{-0.0cm}
\caption{Projected data for the Sivers function vs $x$ for different transverse 
momentum bins (left panel) and vs $p_T$ for different $x$ bins. }
\label{sivers}    
\end{figure}
Similar to the extraction of GPDs, TMD studies will greatly benefit from the higher energy and  
luminosity available at 12 GeV. A comprehensive program is in preparation at JLab to study the 
new quark distribution functions. The main focus is on the measurement of SIDIS processes in 
the full 4-D phase space available in $Q^2$, $x$, $z$ and $p_T$. Fig.~\ref{sidis} 
shows the phase space coverage and statistical precision expected for semi-inclusive $\pi^+$ production
proposed  in~\cite{e12-07-107}.  Examples of kinematics coverage and statistical accuracy are presented in Fig.~\ref{fig:Boer-Mulders}. 

Several experiments will measure the Mulders-Kotzinian asymmetry on longitudinally polarized 
proton target~\cite{e12-07-107,e12-09-009} and on $^3He$ target~\cite{e12-11-007}.  
Figure~\ref{sidis-kaon-AUL} shows an example of the expected asymmetry and statistical precision for charged
kaon production. Several experiments have been proposed to study the Sivers asymmetries on 
polarized $^3$He~\cite{e12-10-006} and on polarized hydrogen~\cite{c12-11-111,c12-11-108}. 
Examples of kinematic dependences are shown in Fig.~\ref{sivers}. 
Although pions have been the main focus of SIDIS experiment in the past, the K$^+$ and
 K$^-$ channels have recently emerged as of high interest as well. Hermes results show 
 unexpectedly large Boer-Mulders asymmetries for kaons compared to pions, and the 
 opposite signs for K$^-$ and $\pi^-$.  
 Several experiments~\cite{e12-09-007,e12-09-008,e12-09-018,e12-09-009,c12-11-111} 
 will measure semi-inclusive processes for kaon productions. With the 
 excellent particle identification and high luminosity expected for {\tt CLAS12}, these puzzling
 issues can be addressed efficiently. Figure~\ref{sidis-kaon-AUL}  shows the 
 statistical precision for K$^+$ and K$^-$ production on longitudinally polarized hydrogen 
 in one kinematic bin in $Q^2$, $z$, and $p_T$.  
 
 The experimental effort will produce high quality asymmetry data in a large range of the 4-dimensional 
 space of $Q^2,~x,~z, ~p_T$. The challenge will be to extract from these large amounts of data the TMDs
 that encode the quark's intrinsic spin-momentum correlations. 

\begin{figure}[t]
\resizebox{1.\textwidth}{!}{%
  \includegraphics{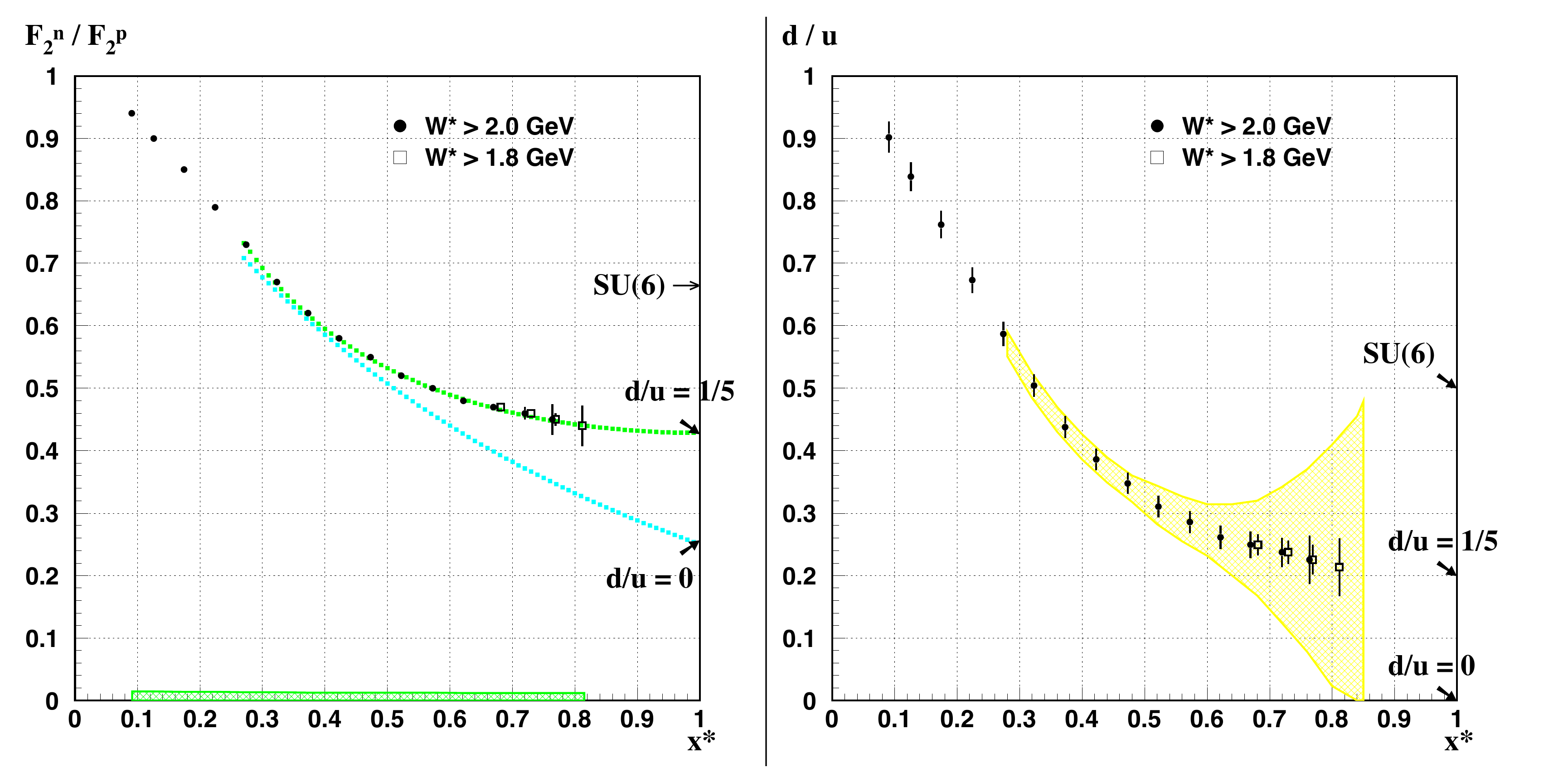}}
\caption{Projected data for the ratio $F^n_2/F^p_2$ (left) and $d/u$ (right) for 
11 GeV beam energy~\cite{e12-10-113}. The error bars in the right panel 
contain both statistical and 
systematic uncertainties. The yellow area shows the uncertainty of current data due to
poorly known nuclear corrections.}
\label{fig:F2nF2p}    
\end{figure}

\begin{figure}[t]
\resizebox{0.5\textwidth}{!}{%
  \includegraphics{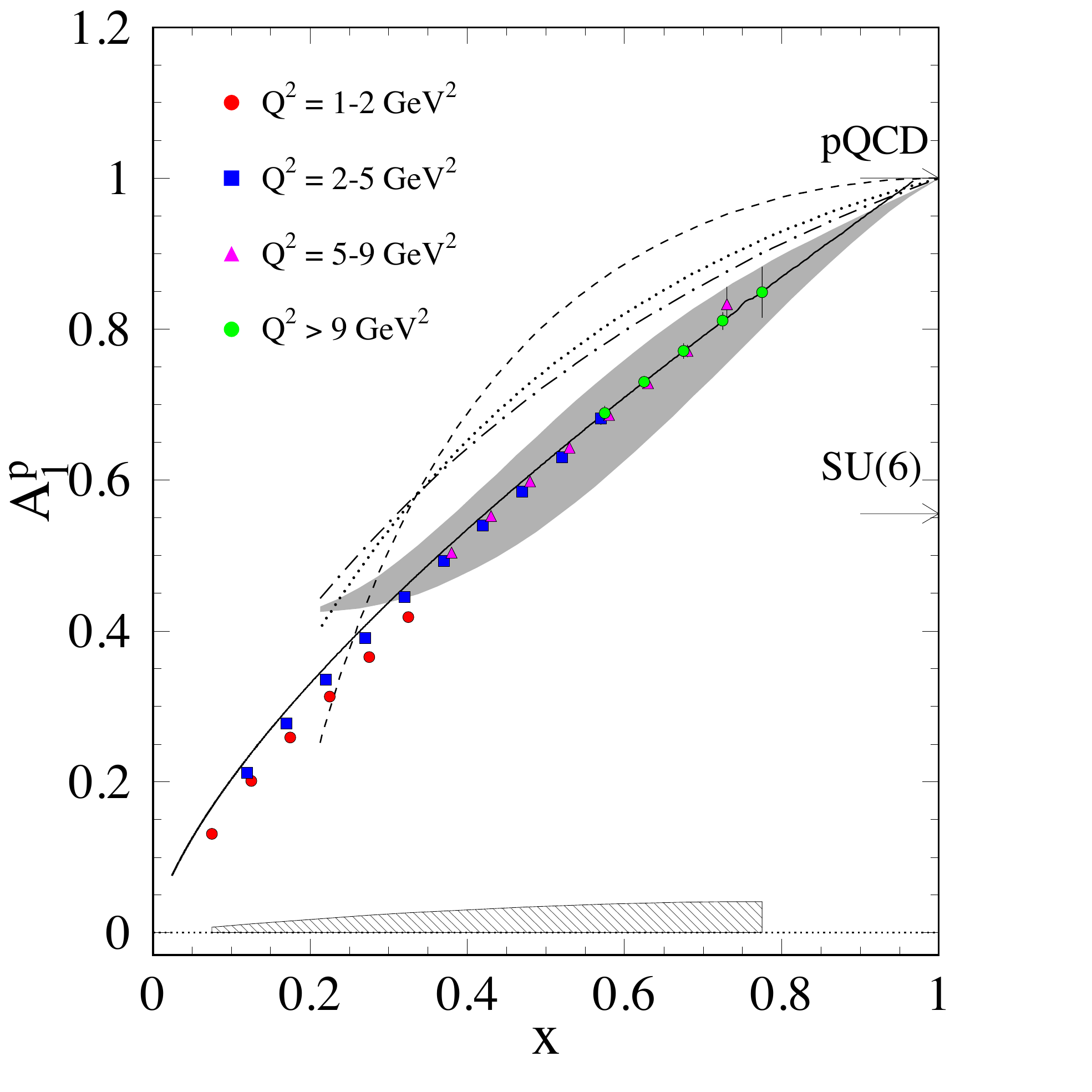} }
  \resizebox{0.5\textwidth}{!}{%
  \includegraphics{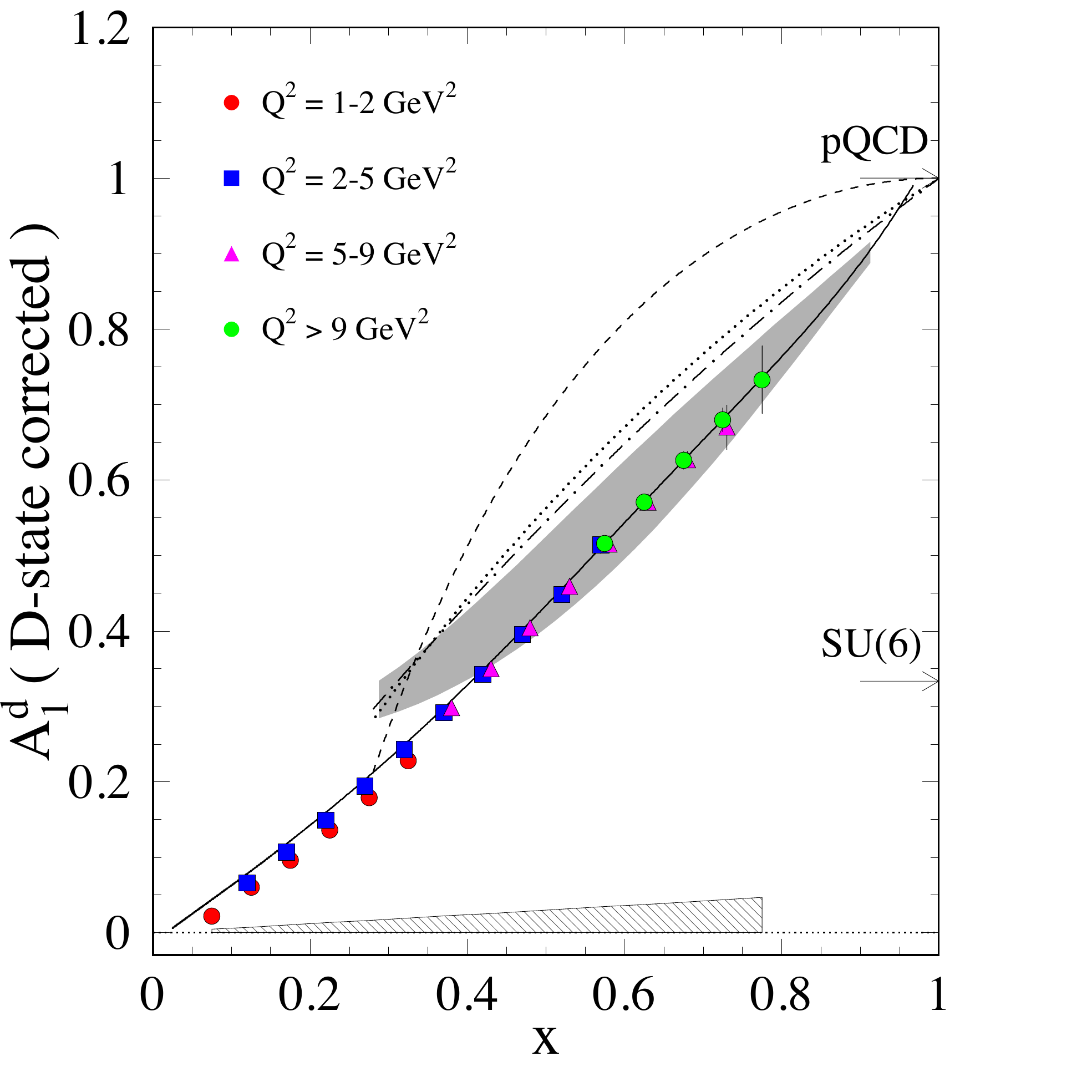}}
\caption{Anticipated results on $A_1^p$.The four different symbols represent four different $Q^2$ ranges.  The statistical uncertainty is given by the error bars while the systematic uncertainty is given by the shaded band. }
\label{fig:A1p}    
\end{figure}
  
\section{Inclusive structure functions and moments}
\label{strf}
Polarized and unpolarized structure functions of the nucleon offer a
unique window into the internal quark longitudinal momentum and helicity 
densities of nucleons.
The study of these structure functions provides insight into the two
defining features of QCD --- asymptotic freedom at small distances,
and confinement and non-perturbative effects at large distance scales.
After more than three decades of measurements at many accelerator
facilities worldwide, a truly impressive amount of data has been
collected, covering several orders of magnitude in both kinematic
variables $x$ and $Q^2$.
However, there are still important regions of the accessible phase space
where data are scarce and have large uncertainties. and where significant
improvements are possible through precise experiments at Jefferson Lab. 

One of the open questions is the behavior of the
structure functions in the extreme kinematic limit $x \rightarrow 1$. 

In this region effects from the virtual sea of quark-antiquark pairs are 
suppressed, making this region simpler to model than the small $x$ region. 
This is also the region
where pQCD can make absolute predictions. However, the large $x$ 
domain is difficult to reach as cross sections are kinematically suppressed, 
the parton distributions are small and final states interactions (partonic or 
hadronic) are large. First steps into the large $x$ domain became possible at energies
of 5-6 GeV \cite{vipuli2006,bosted2007}. 
The interest triggered by these
first results and the clear necessity to extend the program to larger $x$ 
(for recent reviews see:~\cite{christy2011,chen2011})
provided one of the cornerstone of the JLab 12 GeV upgrade physics program.  

\begin{figure}[t]
\resizebox{0.5\textwidth}{!}{%
  \includegraphics{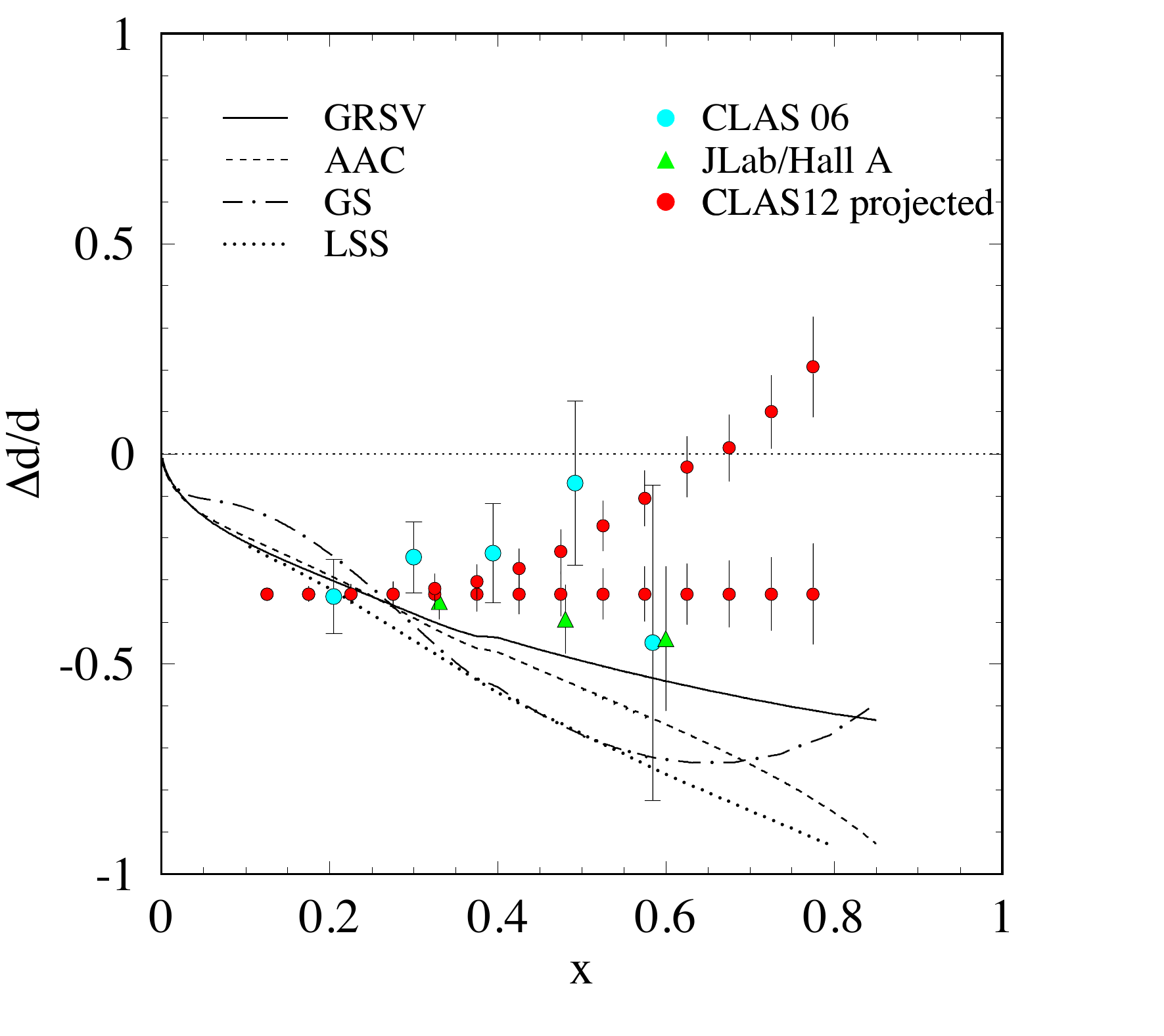}}
  \hspace{-1.5cm}
  \resizebox{0.65\textwidth}{!}{%
  \includegraphics{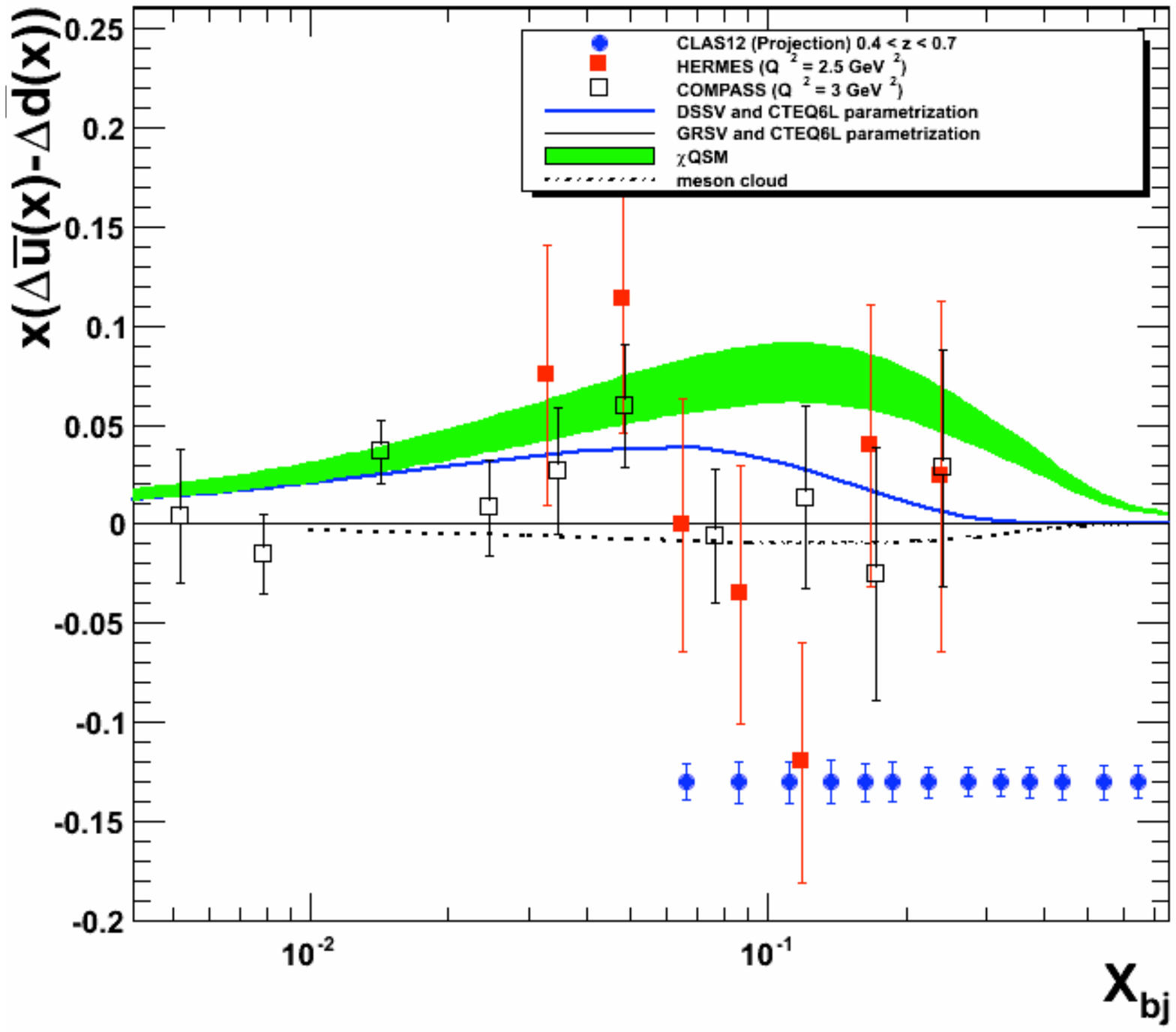}}
\caption{Left panel: Expected results for $(\Delta d+ \Delta\bar{d})/(d+\bar{d})$. The central values of the data are following two arbitrary curves to demonstrate how the two categories of predictions, namely the ones that predict $\Delta d/d$ stays negative (LO and NLO analyses of polarized DIS data: GRSV, LSS, AAC, GS, statistical model, and a quark-hadron duality scenario) and the ones predicting $\Delta d/d \to 1$ when $x \to 1$ (leading order pQCD and a quark-hadron duality scenario). The right panel shows the expected uncertainties for the asymmetry of the polarized sea quarks.}
\label{fig:ddodxpctd}    
\end{figure}
 
\subsection{Valence quark structure and flavor dependence at large $x$.}
\label{valence}

The unpolarized proton structure function $F^p_2(x)$ has been mapped out in a 
large range of $x$ leading to precise knowledge of the quark distribution 
$u(x)$.  The corresponding structure function $F^n_2(x)$ is well 
measured only for $x < 0.5$ as nuclear corrections, using deuterium as 
a target, are large and very uncertain at $x > 0.5$. Corrections for Fermi motion 
are clearly insufficient. 
At JLab, a new technique tested recently has been shown very
effective in reducing the nuclear corrections. The results of the
{\tt BONUS} experiment have recently been published~\cite{baillie2011}. 
The experiment uses a novel radial time-projection-chamber (rTPC) 
with gas-electron-multiplier (GEM) readout as detector for 
the low-energy spectator proton in the reaction $en(p_s)\rightarrow ep_sX$.
Measurement of the spectator proton for momenta as low as 70 MeV/c and at large 
angles minimizes the poorly known nuclear corrections at large $x$. The 
techniques will be used with {\tt CLAS12} to accurately determine 
the ratio $d(x)/u(x)$ to much larger $x$ values~\cite{e12-10-113}. 
Figure~\ref{fig:F2nF2p} shows the projected data for $F^n_2(x)/F^p_2(x)$ and $d(x)/u(x)$. 
A dramatic improvement is projected at large $x$. Another experiment in 
Hall A~\cite{ e12-10-103} measures inclusive scattering on two mirror nuclei $^3He$ 
and $^3H$ for which the nuclear wave functions are very similar. 
The cross section ratio $\sigma(e~{^3He} \to e~X)/\sigma(e~{^3H}\to e~X)$ is 
then largely free of nuclear corrections and the ratio of structure functions 
$F^n_2\//F^p_2$ can be extracted.

\subsection{Spin structure functions and parton distributions}
\label{spin}

The lack of precise data in the valence quark region and especially at very large $x$ values 
is even more obvious for the spin structure function $g_1(x,Q^2)$, both on the proton and neutron. 
New experiments at 12 GeV will significantly improve our knowledge of the basic spin structure function. 
Two experiments will study polarized parton distributions at  $x \le 0.9$ on
polarized protons and polarized neutrons. Using standard detection equipment, a redesigned polarized target adapted to {\tt CLAS12} and 30 (50)~days of running \cite{e12-06-109} on a longitudinally polarized NH$_3$ (ND$_3$) target, high precision results can be achieved as shown in Fig.~\ref{fig:A1p}. Similar coverage is projected using a polarized $^3He$ as a quasi-neutron target in Hall A~\cite{e12-06-110}. These data will discriminate among models in the large-$x$ region. The projected results shown in Fig.~\ref{fig:A1p} are with a $W>2$~GeV constraint. Studies of hadron-parton duality  will tell us if this constraint can be relaxed 
so that spin structure functions may be used for  $x \le 0.9$ in the extraction of parton spin densities from global fits
including these data.
 Results on protons and neutrons targets will allow for
the extraction of the d-quark polarization $(\Delta d+ \Delta\bar{d})/(d+\bar{d})$, and the asymmetry of the polarized sea $x[\Delta\bar{u}(x)-\Delta\bar{d}(x)]$. The accuracy expected from these measurements is shown in Fig.~\ref{fig:ddodxpctd}.

\subsection{Global analysis of polarized parton densities}
\label{fit}
The larger window that will open up over the DIS domain with the 12 GeV upgrade will 
permit more stringent constraints of the parton distributions in global fits to polarized 
structure functions. JLab data at lower energies had already unique impact
at large $x$. The improvement from the 12-GeV upgrade is 
also significant at low and moderate $x$, noticeably for the polarized gluon 
distribution $\Delta G$. To demonstrate the precision 
achievable with the expected {\tt CLAS12} data, we have plotted in 
Fig.~\ref{fig:pDelta_G} the expected impact of future JLab data at 12 GeV on the 
next-to-leading order QCD analyses of the polarized gluon distribution~\cite{LSS2007}.  
A dramatic 
improvement can be achieved with the expected data from experiment 
E12-06-109~\cite{e12-06-109}.  The data will not only 
reduce the error band on $\Delta G$, but will likely allow a more detailed 
modeling of its $x$-dependence. Significant improvements are expected for the quark 
distributions as well, especially for the polarized $s$ quark density.

\subsection{Moments of spin structure functions}
\label{moments}
Moments of structure functions are related to the nucleon static properties 
by sum rules. Inclusive deep inelastic scattering data at JLab have permitted 
evaluation of the moments 
at low and intermediate $Q^2$~\cite{fatemi2003,yun2003,chen2004,vipuli2006,prok2009}.  
With a maximum beam energy of 6~GeV, however, the measured fractional 
strength of the moments becomes rather limited for $Q^2$ greater than a few GeV$^2$. The 
12-GeV upgrade extends this range to much higher $Q^2$. 
\begin{figure}
\resizebox{0.8\textwidth}{!}{%
  \includegraphics{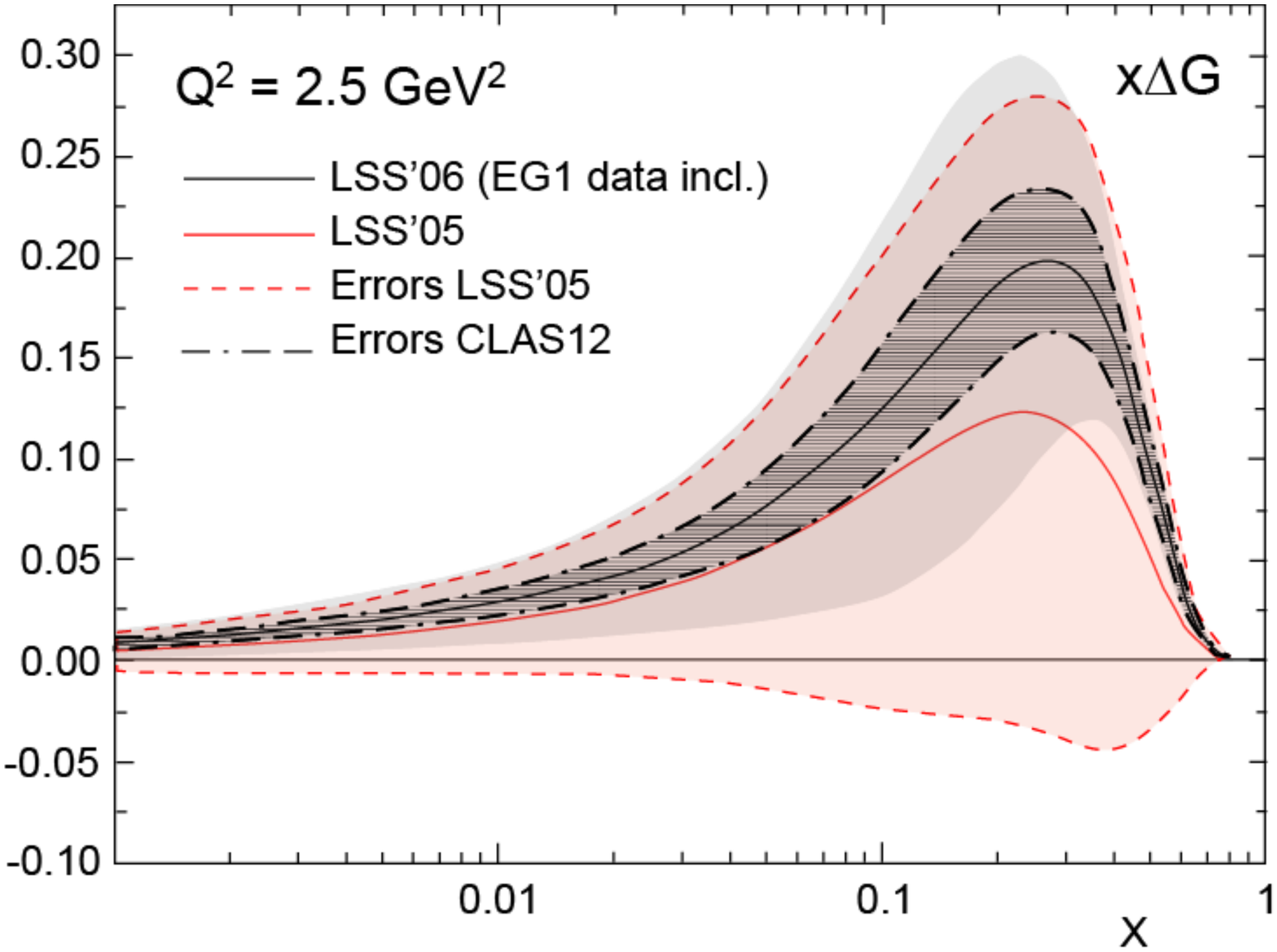}}
\caption{Expected uncertainties for $x\Delta{G}$. The black solid curve shows the central value of the present analysis that includes CLAS EG1 data. The dashed-dotted lines give the error band when the expected {\tt CLAS12} data are included in the LSS QCD analysis.}
\label{fig:pDelta_G}    
\hspace{1.5cm}
\resizebox{0.7\textwidth}{!}{%
  \includegraphics{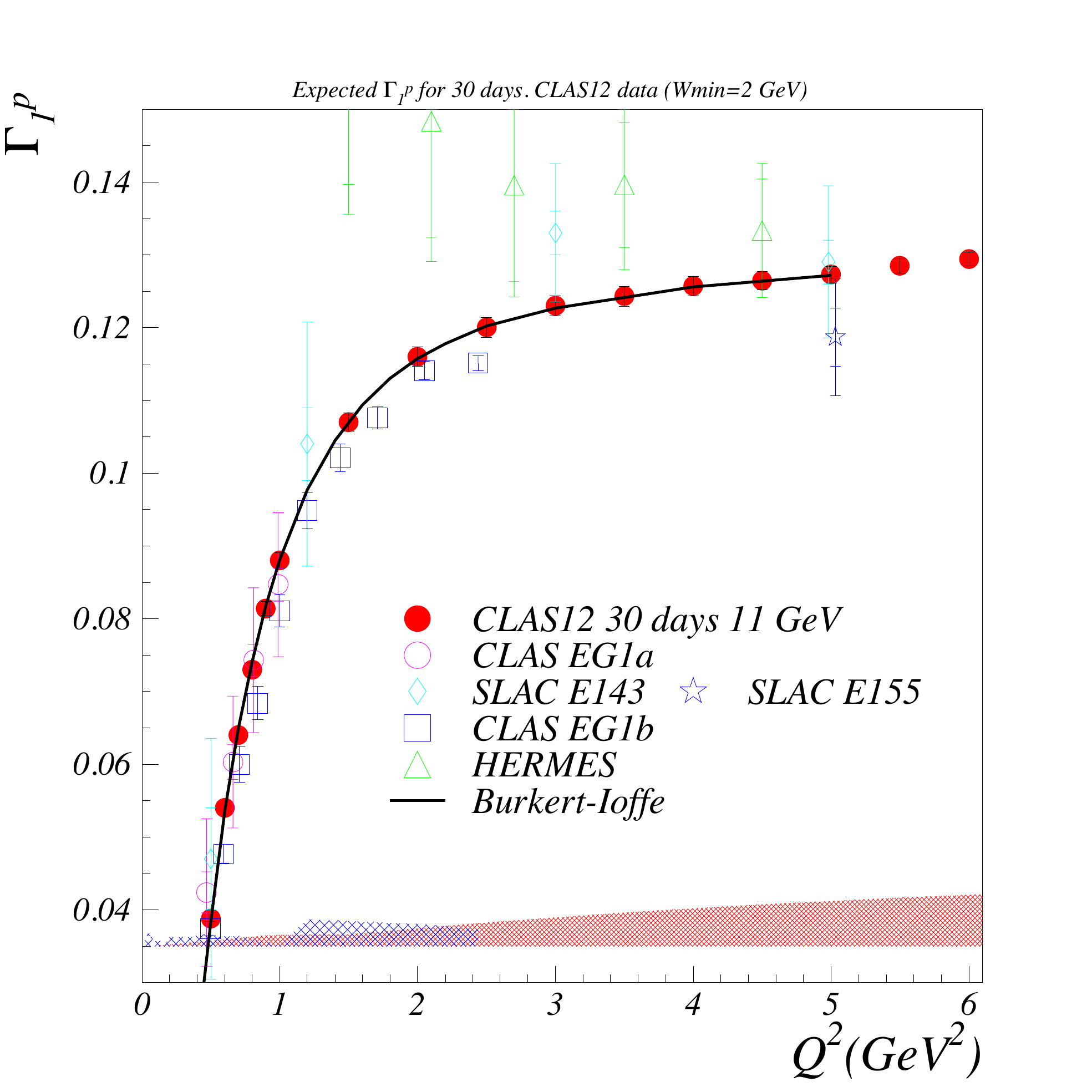}}
\caption{Expected precision on $\Gamma_1^p$ for {\tt CLAS12}
and 30~days of running.  {\tt CLAS} EG1a~\cite{fatemi2003,yun2003} data 
and preliminary results from EG1b are shown for comparison.  The data and 
systematic uncertainties include estimates of the unmeasured DIS 
contribution.  HERMES~\cite{Airapetian:2002wd} data, and E143~\cite{Abe:1998wq} 
and E155 data~\cite{Anthony:2000fn} from SLAC are also shown (including DIS 
contribution estimates).  The solid line is a model from \cite{Burkert:1992tg,Burkert:1993ya}.}
\label{fig:Gamma1-expect}    
\end{figure}
At sufficiently large $Q^2$, the Bjorken sum rule relates the integral 
$\Gamma_1^{p-n}=\int(g_1^p- g_1^n)dx$ 
to the nucleon axial charge~\cite{Bjorken:1966jh}. 
Figure~\ref{fig:Gamma1-expect} shows the expected precision on $\Gamma_1^p$. Published 
results and preliminary results from CLAS are also displayed for comparison. 
The hatched blue band corresponds to the systematic uncertainty 
on the CLAS EG1b data points. The red band indicates the estimated systematic 
uncertainty from {\tt CLAS12}.  The systematic uncertainties for EG1 and 
{\tt CLAS12} include the estimated uncertainty on the unmeasured DIS part 
estimated using a frequently used model~\cite{Thomas:2000pf}.  As 
can be seen, moments can be measured up to $Q^2$=6~GeV$^2$ with a statistical 
accuracy improved several fold over that of the existing world data.

Finally, moments in the low ($\simeq$ 0.5 GeV$^2$) to moderate 
($\simeq$4~GeV$^2$) $Q^2$ range enable us to extract higher-twist parameters,
which represent correlations between quarks in the nucleon. This extraction 
can be done by studying the $Q^2$ evolution of the first moments~\cite{Osipenko2005,Chen:2005td}.
Higher twists have been consistently found to have, overall, a surprisingly 
smaller effect than expected.  Going to lower $Q^2$ enhances the higher-twist 
effects but makes it harder to disentangle a high twist from the yet higher 
ones.  Furthermore, the uncertainty on $\alpha _s$ becomes prohibitive at low 
$Q^2$.  Hence, higher twists turn out to be hard to measure, even at the 
present JLab energies.  Adding data at higher $Q^2$ to the present JLab data set 
removes the issues of disentangling higher twists from each other and of the 
$\alpha _s$ uncertainty.  The smallness of higher twists, however, requires 
statistically precise measurements with small point-to-point correlated 
systematic uncertainties.  Such precision at moderate $Q^2$ has not been 
achieved by the experiments done at high energy accelerators, while JLab at 
12~GeV presents the opportunity to reach it considering the expected 
statistical and systematic uncertainties of the new experiments.  The total 
point-to-point uncorrelated uncertainty on the twist-4 term for the Bjorken 
sum, $f_2^{p-n}$, decreases by a factor of 5-6 compared to results obtained in
Ref.~\cite{Deur:2004ti}. 
\section{Elastic and resonance transition electromagnetic form factors at short distances} 
\label{ff}

\subsection{Nucleon elastic form factors}
\begin{figure}
\resizebox{0.9\textwidth}{!}{%
  \includegraphics{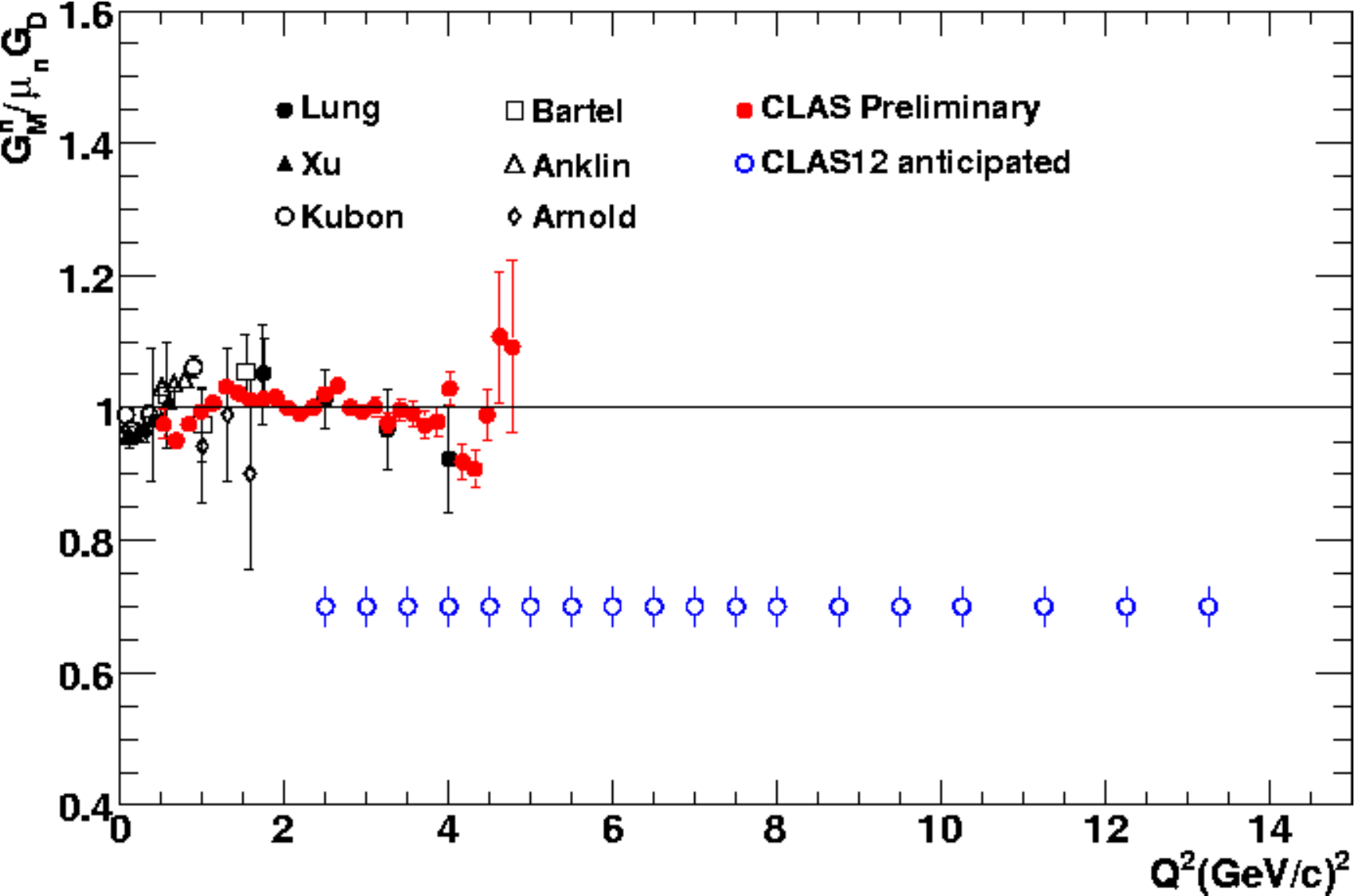}}
\caption{The magnetic form factor for the neutron. The existing data, and projected 
uncertainties at 12 GeV with {\tt CLAS12} (blue open circles).}
\label{fig:gmn}    
\vspace{-0.75cm}
  \resizebox{1.0\textwidth}{!}{%
  \includegraphics{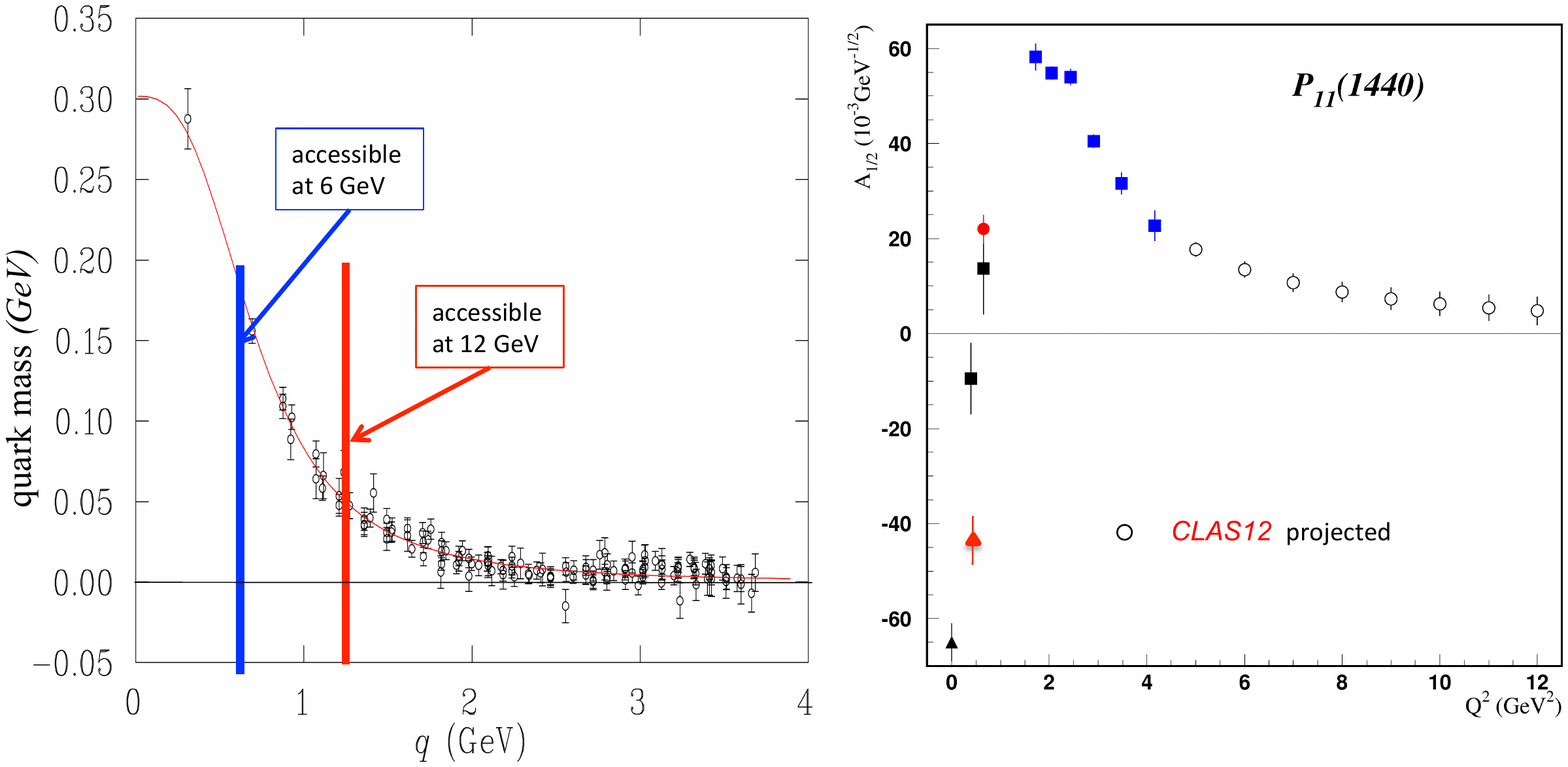}}
  \vspace{-2.5cm}
\caption{Left panel: Evolution of the quark mass with momentum transfer. The region left of the blue line is accessible at 6 GeV beam energy, the region left of the red line can be accessed at 12GeV in N* excitations with $Q^2 \le 12$GeV$^2$. Right panel: Published and projected electro-coupling amplitudes for the Roper resonance with a 12GeV electron beam.  }
\label{quark-mass}    
\end{figure}

The most basic observables that reflect the composite nature of the 
nucleon are its electromagnetic form factors. 
The electric and magnetic form factors characterize the distributions of 
charge and magnetization in the nucleon as a function of the spatial resolving 
power.  Further, these quantities can be described and related to other 
observables through the generalized parton distributions discussed 
in section~\ref{gpds}.

Measurements of the elastic form factors continues will remain an important aspect 
of the physics program at 12 GeV, and will be part of the program in several 
experiments at JLab~\cite{e12-07-104,e12-07-109,e12-09-019,e12-09-016,e12-07-108}. 
The magnetic form factor of the 
neutron requires special experimental setups for neutron detection and the in-situ neutron efficiency 
calibration measurement, for which {\tt CLAS12} is well suited. 
Figure ~\ref{fig:gmn} shows the existing data as well as the extension 
in $Q^2$ projected for the 12 GeV program. 

Nucleon ground and excited states represent different eigenstates of the Hamiltonian, 
therefore to understand the interactions underlying nucleon formation from 
fundamental constituents, the structure of both the ground state and the excited 
states must be studied. 

\subsection{Nucleon resonance transition form factors}
 \begin{figure}[t]
 \hspace{-1cm}
\resizebox{1.1\textwidth}{!}{%
  \includegraphics{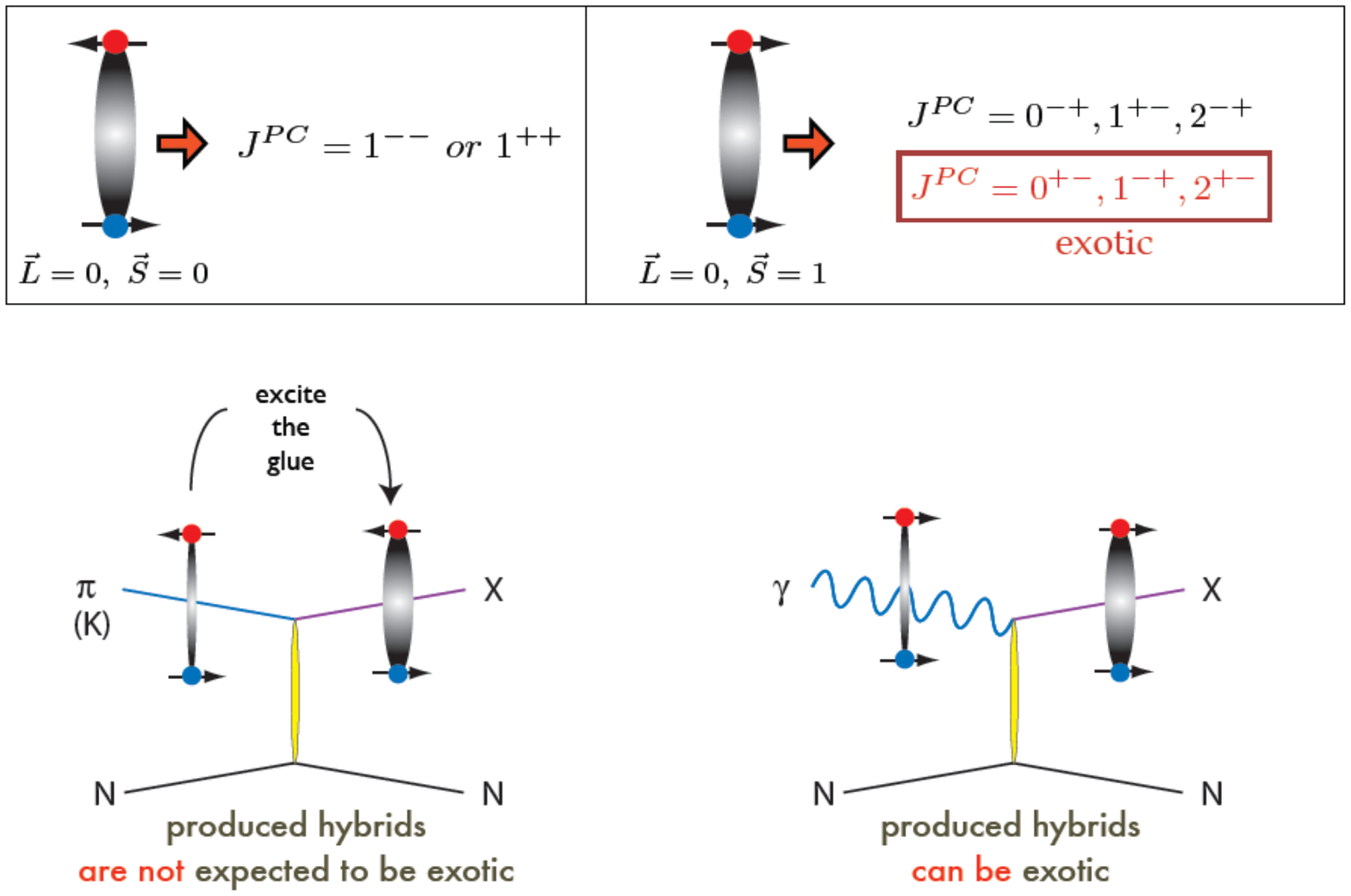}}
  \vspace{-1cm}
\caption{Possible production mechanisms of mesons with exotic quantum numbers with photon beams. In this 
picture mesons with exotic quantum numbers are preferentially produced in reactions with real photons.}
\label{Photoproduction-hybrid-mesons}    
\end{figure}

\begin{figure}[t]
\hspace{-2cm}\resizebox{1.2\textwidth}{!}{%
  \includegraphics{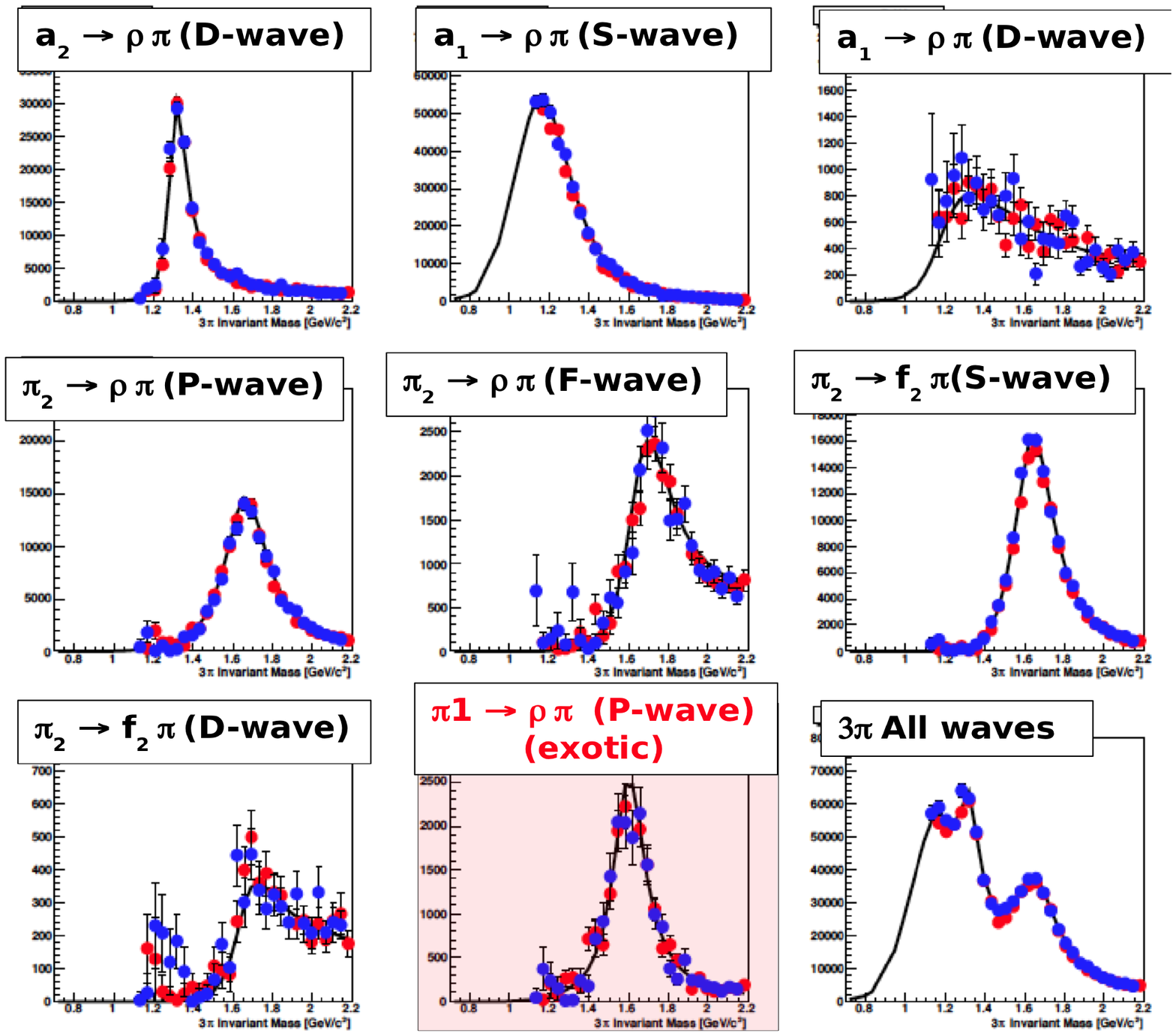} }
\caption{Events with 3-pions in the final state are generated in specific partial waves (solid lines) including one with exotic \protect $J^{PC}$ quantum numbers, and then tracked through the {\tt CLAS12} detector system and then reconstructed (points with error bars). They reconstructed events retain their initial partial wave contents. }
\label{pwa_clas12} 
\end{figure}

The nucleon resonance transition form factors reveal the nature of the excited states
 and encode the transition charge and current densities on the light cone.   
The $N^*$ program at JLab (for recent overviews see:~\cite{aznauryan2011,aznauryan2011a}) 
has already generated results for the transition form factors at $Q^2$ up to 6 GeV$^2$ for the 
$\Delta(1232)$, and up to 7.5~GeV$^2$ for the $N(1535)S_{11}$. 
The latest results~\cite{aznauryan2008,aznauryan2009} 
on the transition form factors of the Roper resonance $N(1440)P_{11}$ for $Q^2$
 up to 4.5~GeV$^2$, have demonstrated the sensitivity to the degrees of freedom 
 that are effective in the excitation of particular states. The JLab energy upgrade will 
 allow us to probe resonance excitations at much higher $Q^2$~\cite{e12-09-103}, 
 where the elementary quarks degrees of freedom may become evident 
 in the resonance formation 
through the approach to scaling as constituents quarks are stripped of the 
gluon cloud and approach their asymptotic mass values. 
Resonance transition form factors are sensitive to the effective quark mass. 
Figure~\ref{quark-mass} shows the projected $Q^2$ dependence 
of the $A_{1/2}$ transition amplitude for the $N(1440)P_{11}$ resonance obtained
 from single pion production. Resonances at higher mass tend to decouple from the 
 single pion channel. The transition form factors of those excited states  
 can be efficiently measured in double-pion processes~\cite{ripani2003} such 
 as $ep\rightarrow ep\pi^+\pi^-$.    

\section{Gluonic Hadron spectroscopy.}
\label{spectroscopy}
The presence of gluons has been established in high energy $e^+e^-$ collisions where the reaction 
$e^+e^-\to 3~$jets was observed. This process is predicted in perturbative QCD and is quantitatively 
described as the result of hard gluon bremsstrahlung from one of the high energy quarks in the annihilation process $e^++e^- \to q + \bar{q} + g$. One of the open problems in hadron physics is if gluons (or the "glue") actively participate processes at  low energy, such as the excitations of hadrons. The most promising way of establishing the effect of the "glue" in spectroscopy 
is the study of the meson spectrum. 
  
\subsection{Hybrid mesons} 
Our understanding of how quarks form mesons has evolved within QCD and we expect a rich spectrum of mesons that takes into account not only the quark degrees of freedom, but also the gluonic degrees of freedom. Excitations of the gluonic field binding the quarks can give rise to excitations of the glue (hybrid mesons).  A picture of these hybrid mesons is one where these particles are excitations of a gluonic flux tube that forms between the quark and antiquark. Particularly interesting is that many of these hybrid mesons are expected to have exotic quantum numbers $J^{PC} = 0^{+-},~1^{-+}, ~2^{+-}$ that cannot be achieved in simple $q \bar{q}$ systems. In the flux tube model these states do not mix with conventional meson states which simplifies dramatically the search for hybrid mesons. The isolation of mesons with exotic quantum numbers provides strong evidence for gluonic excitations, however they do not uniquely identify 
hybrid mesons. Other complex configurations, such as 4-quark states ($q\bar{q}q\bar{q}$ may also have same quantum numbers. For that reason it is essential to establish not only the existence of one or two states, but as many states as possible to characterize the systematics of the spectrum. The level splitting between the ground state flux tube and the first excited transverse modes is expected to be about 1 GeV/c$^2$, and lattice QCD calculations indicate the lightest exotic hybrid (the $J^{PC} = 1^{-+}$) has a mass of about $1.9\pm 0.2$GeV. 

There are tantalizing suggestions, mainly from experiments using beams of $\pi$ mesons, 
that exotic hybrid mesons do exist. The evidence is by no means clear cut, owing in part, to the apparently small production rates for these states in the decay channels examined. It is safe to conclude that the extensive data collected to date with ¹ probes have not uncovered the hybrid meson spectrum. Based on models, such as the flux-tube model, we expect the production of hybrid mesons in photon induced reactions to be comparable to the production of normal mesons.
Photoproduction of mesons using a $\approx 9$ GeV, linearly polarized photon beam provides a unique opportunity to search for exotic hybrids. Existing data are extremely limited for charged final states, and no data exist for multi-neutral final states. To carry out such a search, the GlueX~\cite{e12-06-102}  program in Hall D will look at many different final states involving both charged particles and photons, but particular emphasis will be placed on those reactions that have 3 or more pions in the final state. The discovery potential for GlueX comes first from the very high statistics based on $10^7$ tagged photons per second on target, which will exceed existing photoproduction data by several orders of magnitude. The GlueX experimental search in Hall D has a mass reach up to about 2.8 GeV to observe mesons with masses up to 2.5 GeV/c2.  
\begin{table}[b]
\caption{Expected hadron production rates for 1,000 DIS events at 11GeV beam energy.} 
\vspace{-0.5cm}
\hspace{-4.5cm}
 \resizebox{1.5\textwidth}{!}{%
\includegraphics{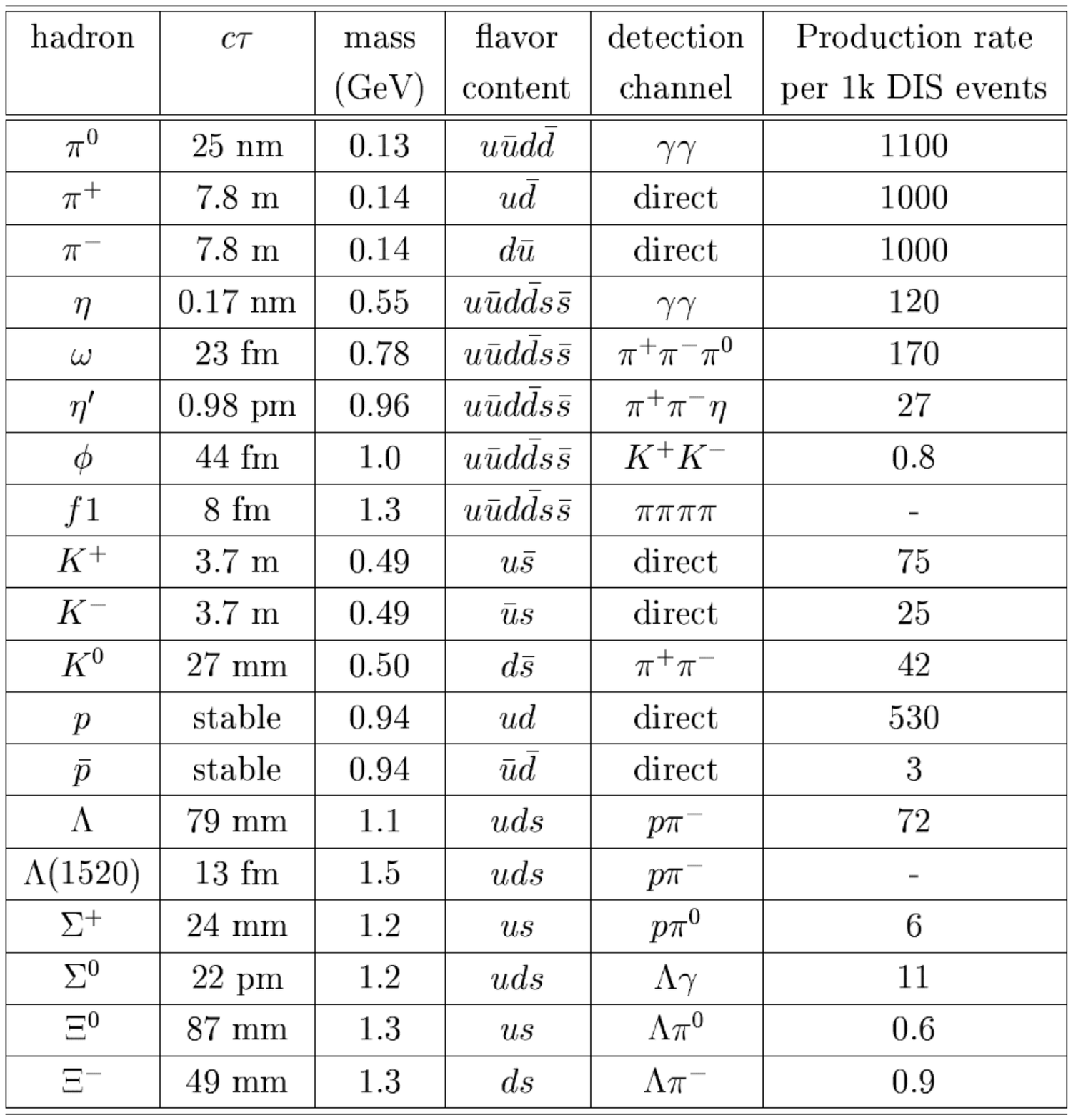}}
\vspace{-1cm}
\label{hadron-species}    
\end{table}

An alternative approach has been taken in Hall B with {\tt CLAS12}, which uses a quasi-real photon beam energy-tagged at very forward electron scattering angles. The Forward Tagger (FT) makes use of a high granularity, high resolution crystal calorimeter to detect the small angle electrons in the energy range of up to 4 GeV~\cite{e12-11-005}.  The difference to photon energy tagging of real photons is that it allows running the experiment at a much higher effective luminosity than is possible when tagging real 
photons. The reason is that real photon tagging requires two independent interactions - the real photon being generated 
in an electromagnetic interaction with the atomic electrons of the radiator by bremsstrahlung, and the real photon interacting hadronically with the proton in the production target. The latter process has a typical probability of about 1\% only. In virtual photon tagging the photon interacts hadronically in the production target, potentially allowing much higher production rates to be achieved in experiments compared to real photo-production. Another aspect is that virtual photons are linearly polarized providing an important degree of freedom that can be utilized in the partial wave analysis of multi-particle final states to reduce ambiguities. 
Figure~\ref{pwa_clas12} shows the results of a partial wave analysis. The events are simulated with different
partial wave content, then tracked through the {\tt CLAS12} acceptance and then subjected to a partial wave analysis. The results of the analysis show that the initial partial wave content is preserved.    
\subsection{Hybrid baryons}
The search for gluonic excitations has so far be limited to mesons, however baryons may also be excited through the their gluon degrees of freedom. Studies of hybrid baryons are equally important to hybrid mesons in the quest to better understand the complex confinement mechanisms in hadrons. However, gluonic baryons $|qqqG>$ and ordinary 3-quark baryons $|qqq>$ have the same quantum number, making it impossible to uniquely separate them from each other. One possibility is to distinguish different types of baryons due to a possible overpopulation 
of states with specific spin-parity $J^P$ assignments in comparison to the number of such states that are predicted in the constituent quark model. The excited glue adds $\approx 1$~GeV to the mass of hybrid baryons, putting the lowest mass hybrid baryons with strangeness $S=0$ into the range from 1.8 to 2~GeV.

\section{Quarks and hadrons in the nuclear medium}
\label{nuclei}

\subsection{Color Transparency}
\label{ct}
Color transparency (CT) is a unique prediction of QCD, and implies that under certain conditions, 
the nuclear medium will allow the transmission of hadrons with reduced absorption. The phenomenon 
of CT is predicted on the quark-gluon basis and is totally unexpected on a hadronic interaction 
picture. the following conditions must be present to observe CT:  the interactions 
must create a small size object (point-like configuration, PLC) that has a small cross section 
when traveling in a hadronic medium, and the distance over which it expands to its full hadronic 
size must be larger than the nucleus size.    
Such conditions require high enough energy transfer to the target where the photon couples to PLCs, 
and the full hadronization occurs outside the nucleus. Color transparency effects have been observed 
at very high energies at hadron machines, but at what momentum transfer CT sets in is still a not fully 
settled question. Small increases in nuclear transparency consistent with theoretical predictions have been observed at JLab with 5-6 GeV electron beams in pion production. The energy doubling of the JLab electron accelerator  to 12 GeV will provide much better conditions where a significantly increased 
transparency should be observable~\cite{e12-06-106,e12-06-107} with high sensitivity, as is shown, e.g. in 
Fig.~\ref{clas12-ct}.   
\begin{figure}
\hspace{-0.3cm}
  \resizebox{0.55\textwidth}{!}{%
  \includegraphics{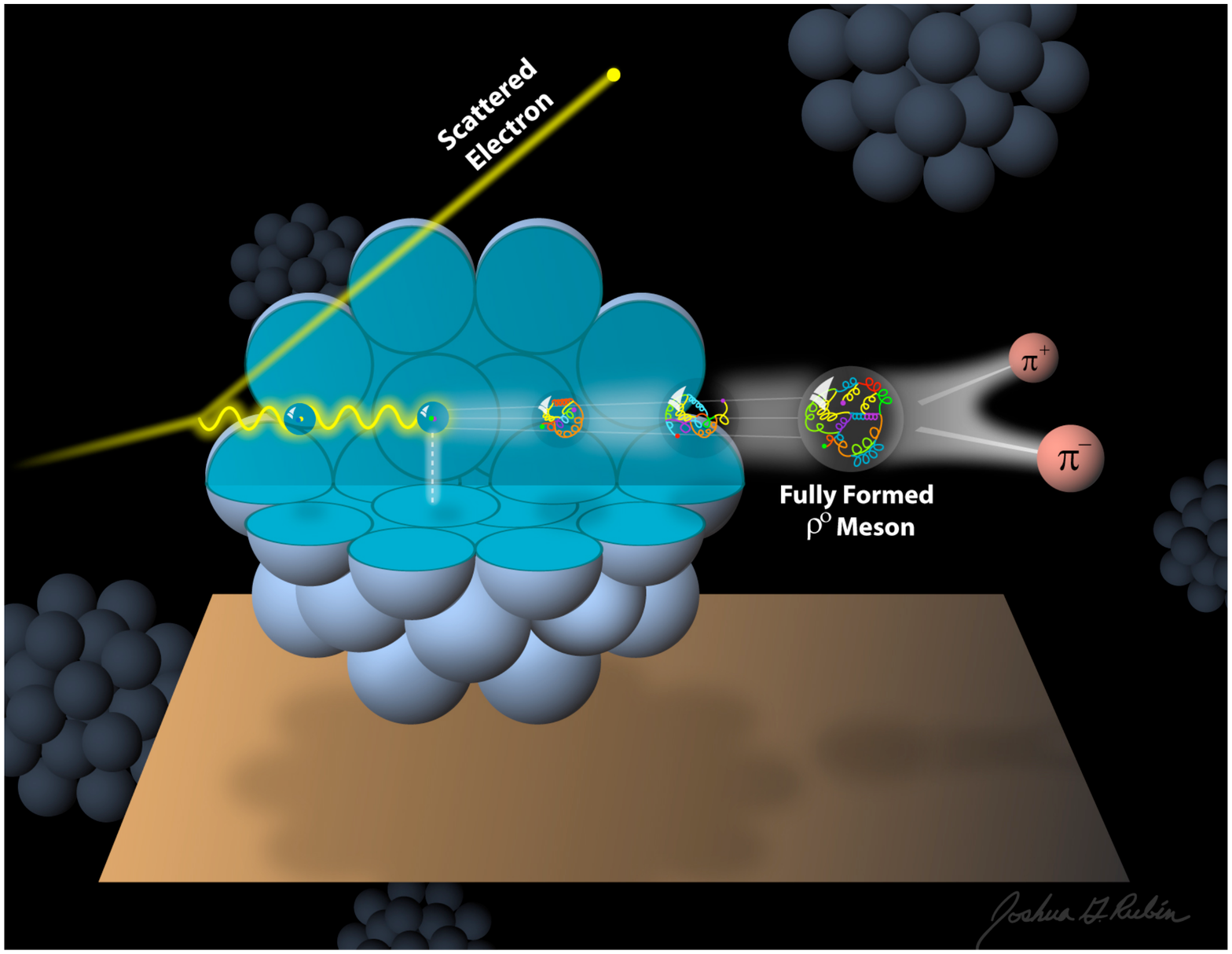}}
  \hspace{-0.8cm}
  \resizebox{0.55\textwidth}{!}{%
  \includegraphics{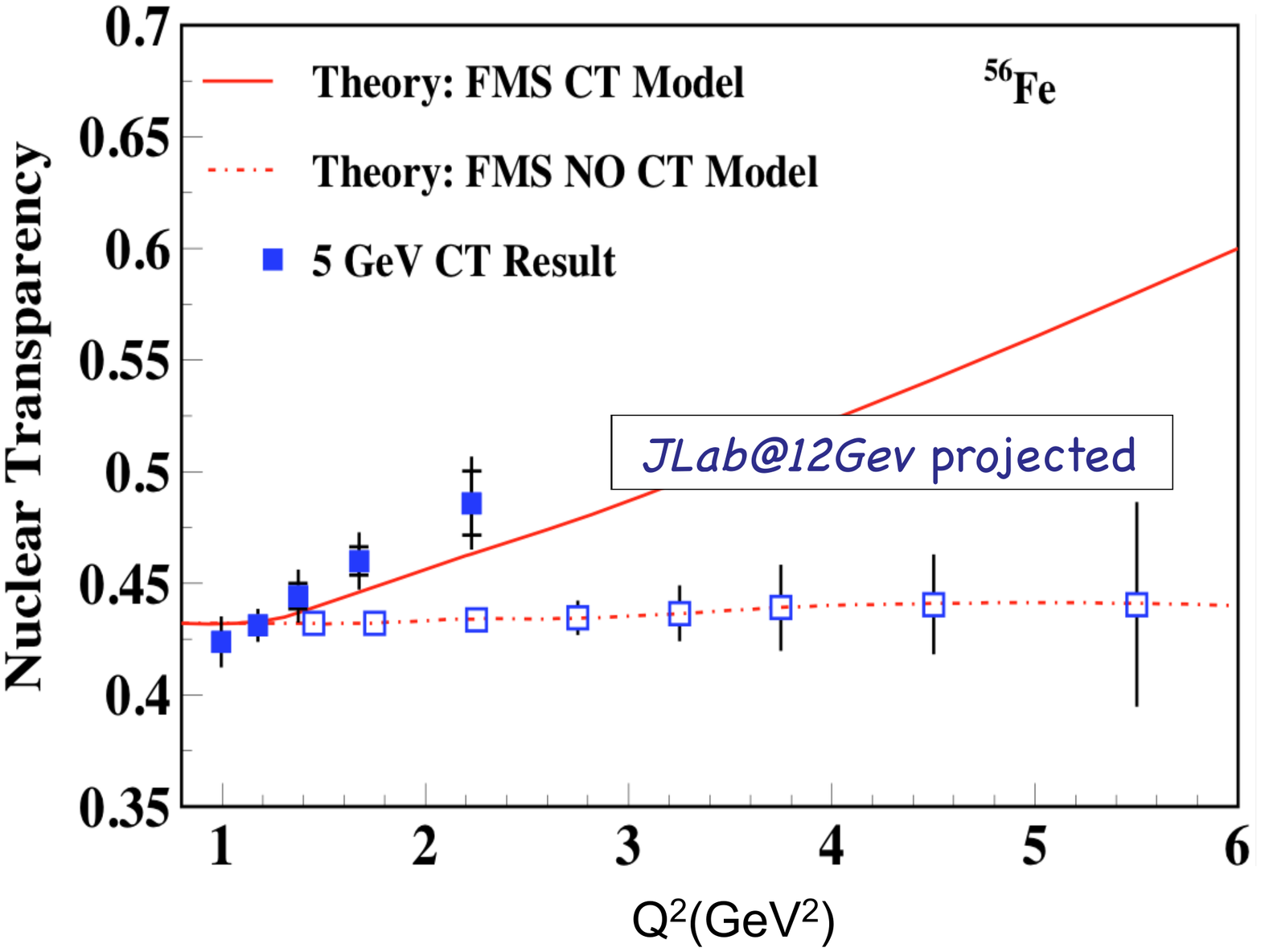}}
\caption{Projected color transparency effects in Fe. The open 
circles represent projected results with {\tt CLAS12} at 12 GeV.}
\label{clas12-ct}
\hspace{-0.5cm}
\vspace{0cm}
  \resizebox{1.1\textwidth}{!}{%
  \includegraphics{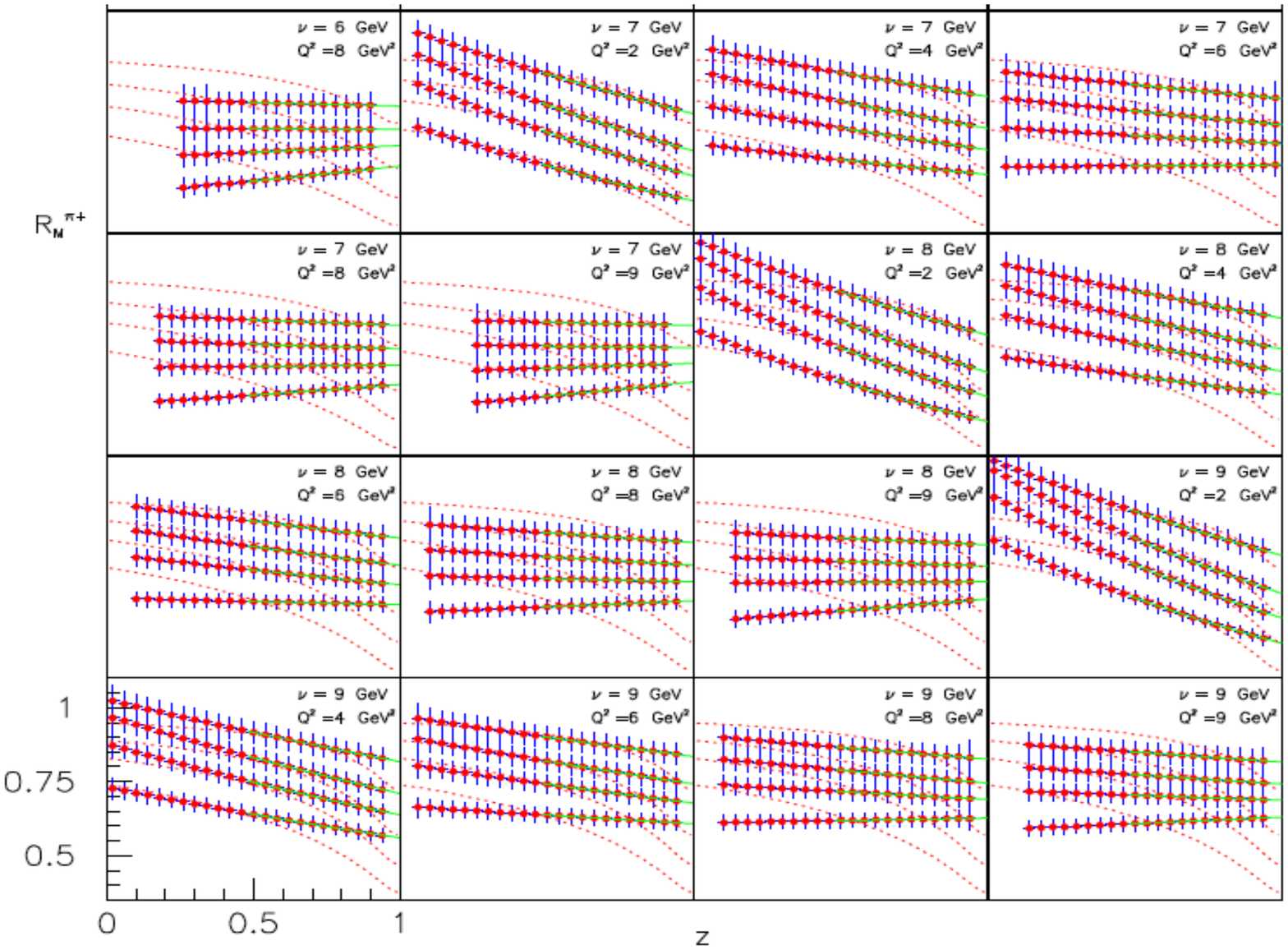}}
\caption{Z dependence of the hadronic multiplicity ration (top to bottom in each plot) for \protect $^{14}N,~^{40}Ar,~^{197}Au$ for 16 bins in
$Q^2,~ \nu$. The solid line is a gluon bremsstrahlung model calculation~\cite{Kopeliovich2004} for $z > 0.5 $ for pions. The dotted lines are parameterizations of HERMES 27 GeV data. } 
\label{hadronization}    
\end{figure}

\subsection{Quark propagation and hadron formation.} 
\label{quark-propagation}
The use of electron beams at 12 GeV allows us to address fundamental questions of how colored quarks struck in the interaction with high energy photons transform into colorless hadrons. Questions that we want to have answered are, how long can a light colored quark remain de-confined? The production time $T_p$ measured this quantity. Because de-confined quarks emit gluons, $T_p$ can be measured via medium-stimulated gluon emission resulting in a broadening of the transverse momentum distribution of the final hadrons. Another important question to address is: How long does it take to form the color field of a hadron? This can be measured by the formation time $T_f^h$. Since hadrons interact strongly with the nuclear medium, $T_f^h$ can be determined by measuring the attenuation of hadrons in the nuclear medium by using nuclei of different sizes.

These question can be addressed by measuring the hadronic multiplicity ratio 
$$R_M^h(z,\nu,p_T^2,Q^2,\phi) = {\{{N_h^{DIS}(z,\nu,p^2_T,Q^2,\phi) \over N_e^{DIS}(\nu,Q^2)}\}_A \over  \{{N_h^{DIS}(z,\nu,p^2_T,Q^2,\phi) \over N_e^{DIS}(\nu,Q^2)}\}_D }$$ 
versus all kinematical quantities. In this expression, $N_h$ is the number of hadrons produced in DIS events and $N^{DIS}_e$ 
is the number of associated DIS electrons. The numerator corresponds to target nucleus $A$, and the denominator corresponds to deuterium. $\nu$ is the energy transferred by the electron, and $z$ is the hadron energy divided by $\nu$. In the QCD-improved parton model, $R^h_M$ is given by the ratios of sums over products of the quark distribution functions with fragmentation
 functions.  This measurement~\cite{e12-06-117} will provide two to three orders of magnitude more data than any previous experiment in this energy range and will include a much larger collection of hadron species. 
 Table~\ref{hadron-species} shows the expected rates for the production of a large number of hadron species. Examples of
 projected data for the multiplicity ratio $R_M^h(z,\nu,p_T^2,Q^2,\phi)$ are shown in Fig.~\ref{hadronization}.

\section{Search for new Physics}
\label{new-physics}
\subsection{Search for deviations from the Standard Model}
The Standard Model (SM) of High Energy Physics has been highly successful in accurately 
predicting a large number of experimental observations. Its predictions have been probed at 
the highest energy accelerators at HERA, the Tevatron at Fermilab, and now at the 
LHC at CERN. So far, no clear deviations from the SM have been identified. 
While most of the searches for physics beyond the SM  
involve very high-energy accelerators where new particle types may be produced in high-energy
collisions, deviations from the SM may also be observed in high precision experiments at 
lower energies. Parity violating electron scattering may be the best 
 tool to study possible SM violating effects in precision measurements of the electro-quark
 couplings that relate to the weak mixing angle $\sin^2\Theta_W$. New physics such as quark 
 compositeness, or new gauge bosons beyond the $Z^0$, may cause small deviations from the 
 SM predictions at relatively low energies that can be revealed in very high precision 
 experiments. With the high luminosity available at JLab such effects  
 maybe be detectable. Two experiments have been proposed to measure with 
 high precision the weak mixing angle $\sin^2\Theta_W$ in parity violating electron-electron 
 scattering (M\/oller scattering)~\cite{e12-09-005}, and in deep inelastic electron scattering~\cite{e12-10-007}. 
 Figure~\ref{weak-mixing} shows the Standard Model prediction for the  energy 
 dependence of the weak mixing angle in comparison with measurements. Previous measurements 
 indicate a possible discrepancy for several of the data points each one at the $\ge 1 \sigma$ level. 
 The new measurements will significantly improve the precision of the experimental  data base. 
\begin{figure}[t]
\vspace{-0.5cm}
\hspace{-0.5cm}
  \resizebox{0.95\textwidth}{!}{%
  \includegraphics{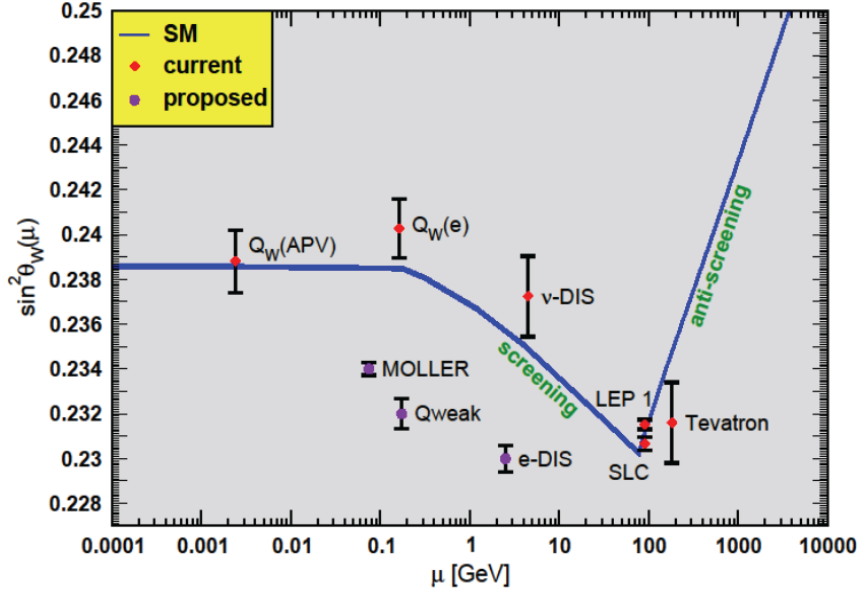}}
\caption{Current knowledge of the weak mixing angle in comparison to the SM predictions . } 
\label{weak-mixing}    
\end{figure}

\begin{figure}[t]
 \resizebox{0.4\textwidth}{!}{%
\includegraphics{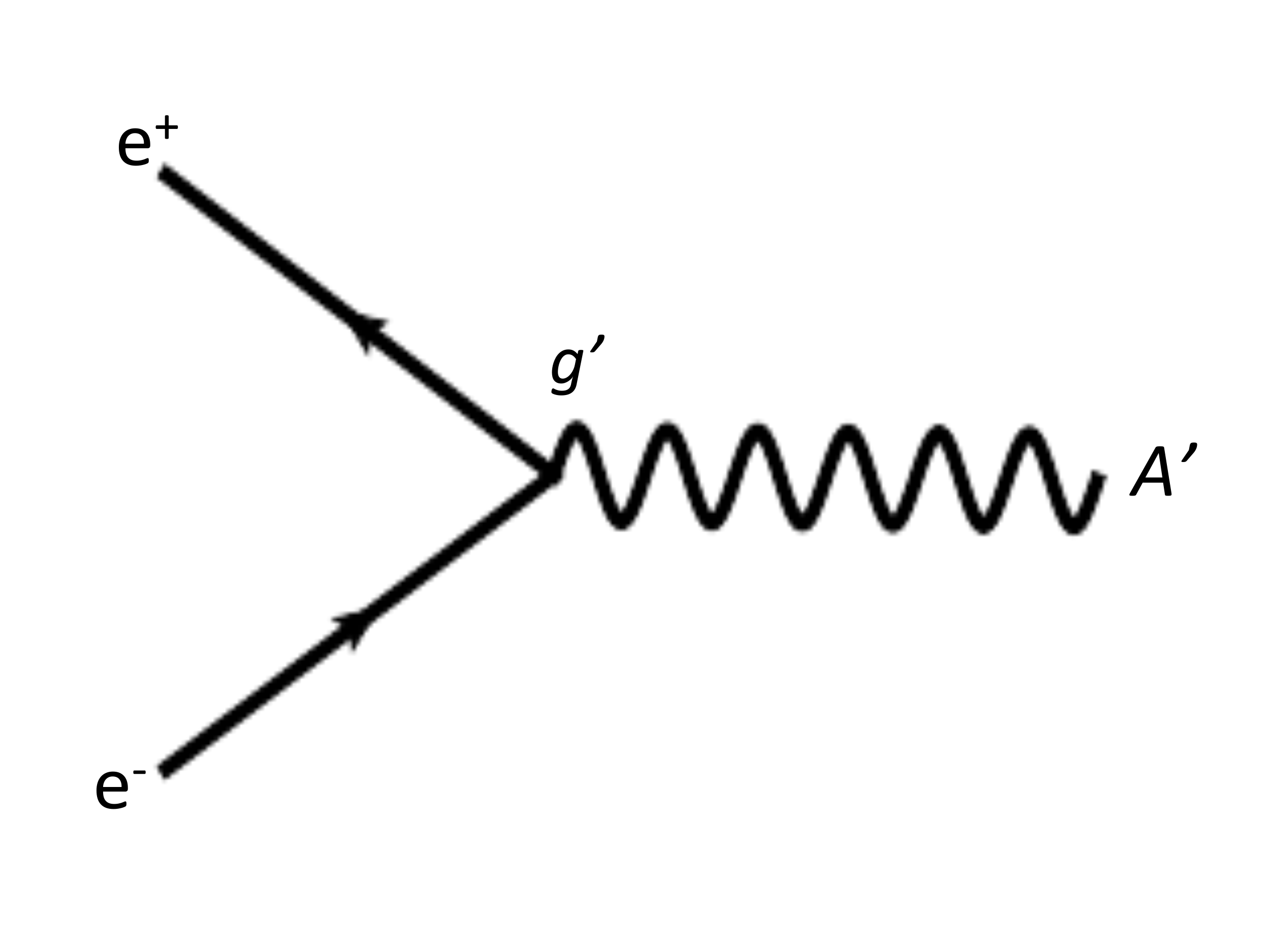}}
 \resizebox{0.4\textwidth}{!}{%
\includegraphics{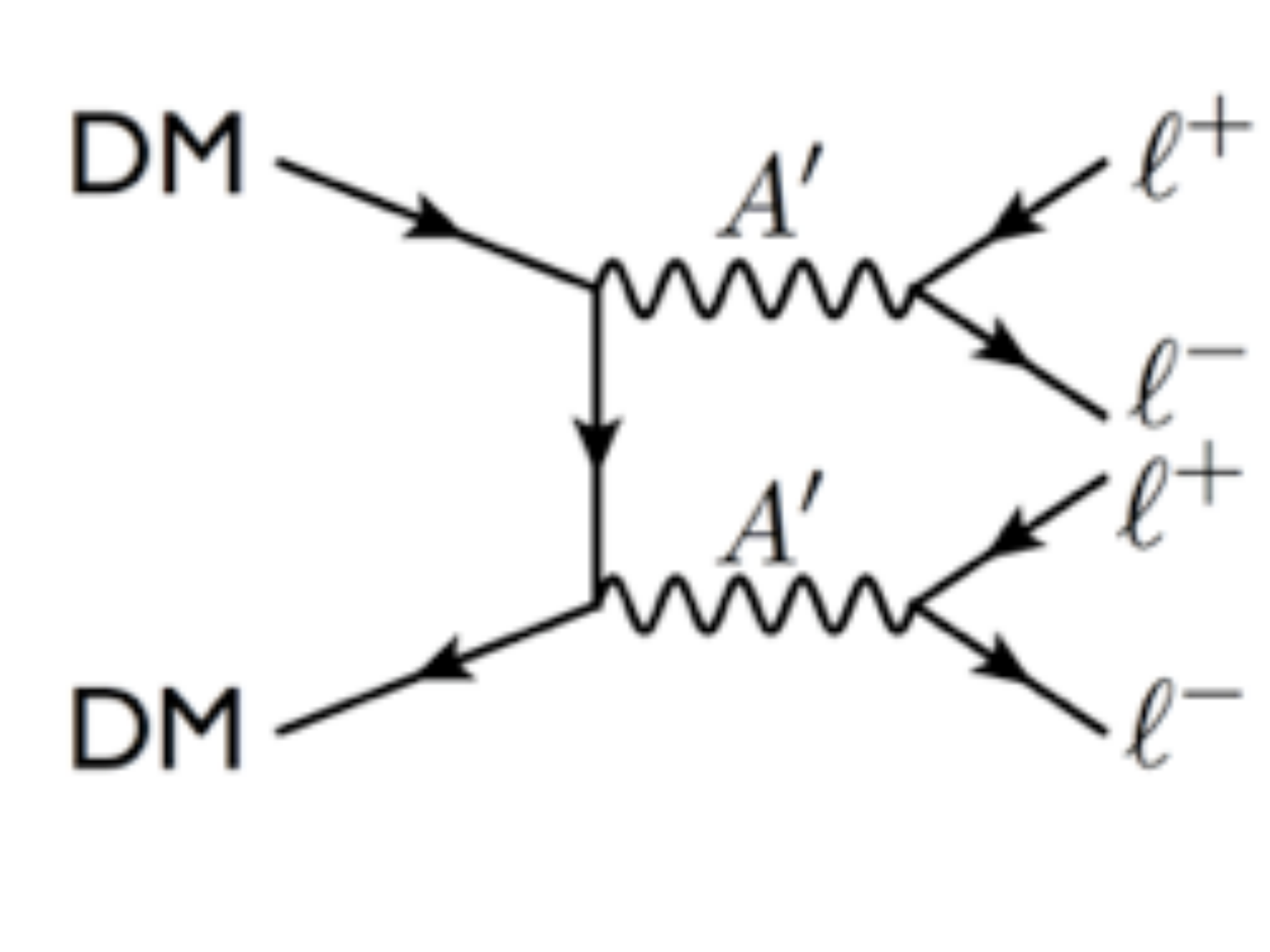}}
\caption{Possible graphs contributing to the excess of positrons (left) and the coupling of the purported $A^\prime$ vector-boson to dark matter (right). }
\label{heavy-photon-graph}    
 \resizebox{1.0\textwidth}{!}{%
\includegraphics{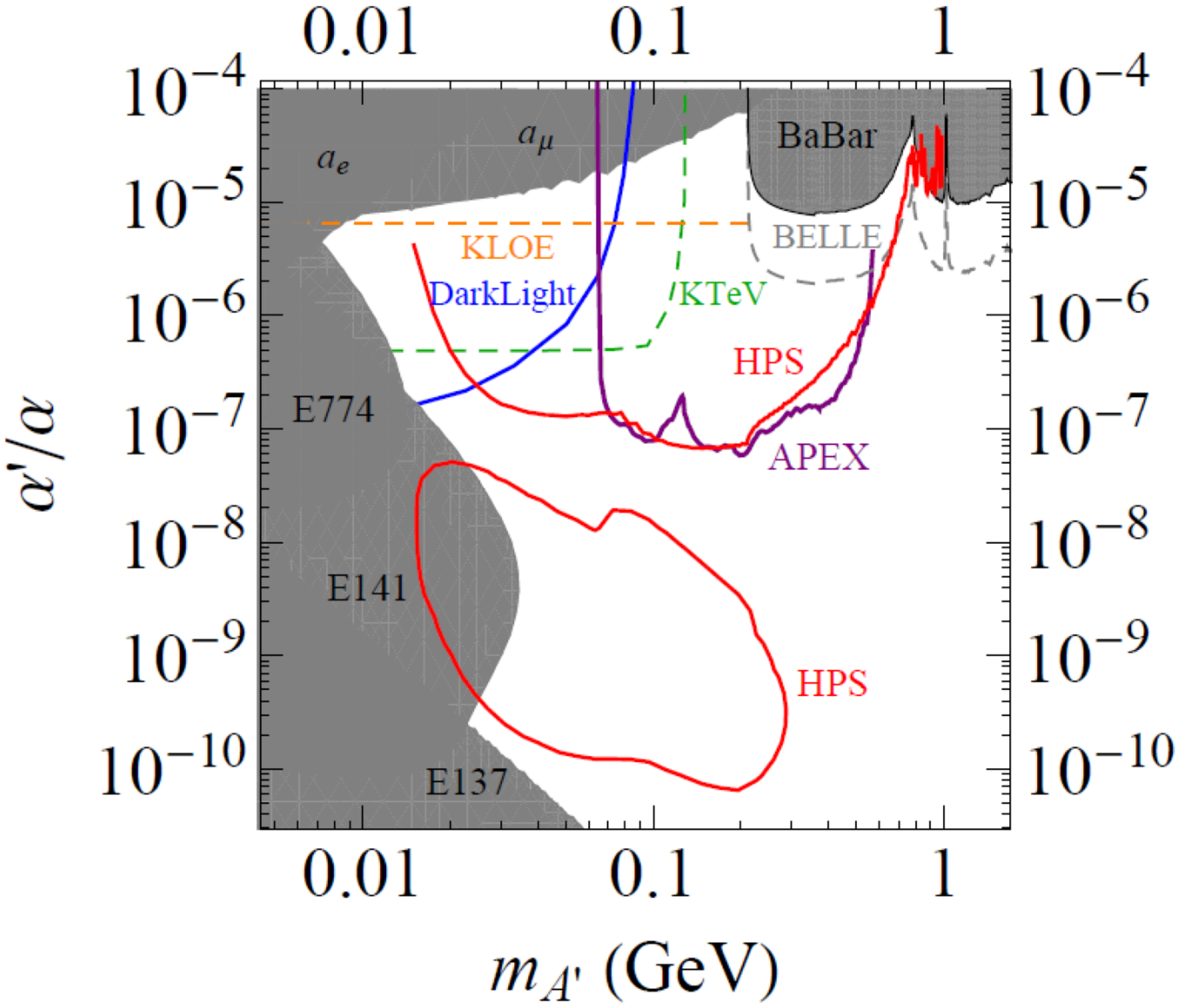}}
\vspace{-1cm}
\caption{Exclusion zones for the ratio of heavy photon to electromagnetic coupling strengths to $e^+e^-$ pairs versus the heavy photon mass. The various lines show the exclusion zones from previous and planned measurements.The proposed and planned measurements at JLab are shown by the lines marked HPS and APEX. The two zones in the upper part are for the HPS and APEX experiment to search for narrow peaks in the mass spectrum of the $e^+e^-$ pair. The zone marked HPS in the lower part includes the search for detached vertices that would indicate the presence of a long-lived $A^\prime$ with weak coupling. }
\label{heavy-photon-exclusion}    
\end{figure}

\subsection{Heavy photon search} 

Recent satellite measurements  have shown a dramatic excess of positrons ranging in 
energy from 10GeV to several hundred 100GeV. One interpretation for this anomaly is the existence of 
a new gauge vector boson that couples very weakly to $e^+e^-$ pairs and
can be produced through interaction with the dark matter in the universe, see Fig.~\ref{heavy-photon-graph}. 
New light vector particles and their associated interactions are ubiquitous in extensions 
of the Standard Model. However, the symmetries of the Standard Model restrict the interaction 
of ordinary matter with such new states. Indeed, most interactions consistent with Standard Model 
gauge symmetries and Lorentz invariance have couplings suppressed by a high mass scale. One 
of the few unsuppressed interactions is the coupling of charged Standard Model particles to a
 new gauge boson $A^\prime$, which is quite poorly constrained for small coupling constants as shown
in Fig.~\ref{heavy-photon-exclusion}. 

Heavy photons mix with the ordinary photon through kinetic mixing, which induces their weak coupling to electrons, $\epsilon  e$, where $\epsilon \approx 10^{-3}$.  Since they couple to electrons, heavy photons are radiated in electron scattering and can subsequently decay into narrow $e^+e^-$ resonances which can be observed above the copious QED trident background. For suitably small couplings, heavy photons travel detectable distances before decaying, providing a second signature. Two experiments have been 
proposed at JLab~\cite{e12-10-009,c12-11-006}  to extend considerably the kinematic zone covered by experiment in the search for a new gauge boson. The APEX experiment in Hall A~\cite{e12-10-009}  uses
high resolution magnetic spectrometers to search for narrow peaks in the $e^+e^-$ mass spectrum. The HPS
experiment in Hall B, in addition to searching for peaks in the mass spectrum also searches for detached
vertices from the decay of a long-lived $A^\prime$. The latter probes a region with much weaker coupling of the
$A^\prime$  to $e^+e^-$ pairs as seen in the bottom part of Fig.~\ref{heavy-photon-exclusion}.  

\section{Conclusions}
\label{summary}

The JLab energy upgrade and the planned new experimental equipment are well 
matched to an exciting scientific program aimed at studies 
of the complex nucleon structure in terms of the newly discovered 
longitudinal and transverse momentum dependent parton distribution 
functions, the GPDs and TMDs. 
They provide fundamentally new insights in the complex multi-dimensional 
quark structure of the nucleon. The exploration of the gluonic meson excitations 
is a complementary effort to study the active role of gluons play in the excitation 
spectrum of hadrons. In addition, the high precision afforded by the 
high luminosity and the large acceptance detectors, and the development of novel techniques to 
measure scattering off nearly free neutrons, will enable the exploration of phase space domains 
with extreme conditions that could not be studied before. 

\acknowledgments 
 
I like to thank my colleagues at Jefferson Lab for many discussions on the exciting science with the 
JLab 12GeV upgrade and for providing me with many of the graphs that I included in this contribution. 
This work was carried out under DOE contract No. DE-AC05-06OR23177.

\end{document}